\documentclass[intlimits,twoside,a4paper]{article}

\usepackage[cp1251]{inputenc}

\usepackage{graphicx}
\usepackage{eso-pic}
\usepackage{xcolor}

\usepackage[eqsecnum]{cmpj3}

\issue{2026}{29}{3}{33801}
\doinumber{10.5488/CMP.29.33801}
\title[Adsorption of LDL on micropatterned brush]%
{Adsorption of the low density lipoproteins on the micropatterned polymer brush: computer simulations
}
\author[J. Ilnytskyi, D. Yaremchuk, O. Komarytsia]
{
J. Ilnytskyi\orcid{0000-0002-1868-5648}\refaddr{label1,label2}\thanks{Corresponding author: \email{iln@icmp.lviv.ua}},
D. Yaremchuk\orcid{0000-0003-2888-5878}\refaddr{label1},
O. Komarytsia\orcid{0000-0002-5822-8281}\refaddr{label3}
}
\addresses{
\addr{label1} Yukhnovskii Institute for Condensed Matter Physics of the National Academy of Sciences of Ukraine, 1~Svientsitskii Str., 79011 Lviv, Ukraine
\addr{label2} Institute of Applied Mathematics and Fundamental Sciences, Lviv Polytechnic National University, 5~Mytropolyta Andreia Str., 79013 Lviv, Ukraine
\addr{label3} Department of Internal Medicine No.~2, Danylo Halytskyi Lviv Medical University, 69 Pekarska Str., 79010 Lviv, Ukraine
}
\date{Received 28 August 2026; accepted 2 September 2026; published 28 September 2026}

\begin{document}

\maketitle

\begin{abstract}
Photorenewable polymer adsorbents, aimed at reducing the low density lipoproteins (LDLs) level, are reported experimentally.  Previously we developed a mesoscale model for this setup with uniformly grafted chains, and examined the role played by chains length and grafting density [\textit{Ilnytskyi, et al., Processes}, 2023, \textbf{11}, 2913]. The analysis is extended here to the cases of bunch-like and grid-like grafting micropatterns. The bunch-like arrangement is the most efficient one displaying low- and high-density peaks for adsorbing effectiveness. The former peak is twice higher compared with uniform grafting, as a consequence of a match between LDLs dimensions and the micropattern pitch. Adsorption isotherms are fitted well by either Langmuir or logistic growth forms, indicating a higher growth rate compared with uniform grafting. The high-density peak exists only for this micropattern and extends the applicability of an adsorbent. At the highest LDL concentration, adsorption is hampered by self-assembly of LDLs into a packed cubic phase.
      
\printkeywords
\end{abstract}

\newcommand{\nm}{\,\textrm{nm}}
\newcommand{\fs}{\,\textrm{fs}}
\newcommand{\ns}{\,\textrm{ns}}
\newcommand{\J}{\,\textrm{J}}
\newcommand{\vvec}[1]{\mathbf{#1}}
\newcommand{\vhat}[1]{\widehat{\mathbf{#1}}}
\newcommand{\Ang}{\,A}

\section{Introduction}\label{sec1}

The excess of the LDL in blood plasma endangers the blocking of arteries and increases risk for cardiovascular diseases \cite{Sacks2003, Sikorski2007, Mundi2017, BasuRay2019, Thierer2019}. The post factum remedy of blocked arteries is their stenting, but a preferable preventive measure to avoid such conditions is a reduction of the LDL concentration in a blood. Besides the medication by statins, ezetimibe, or fibrates, there is another approach, a hemoperfusion therapy, which is gaining more interest. In a course of this therapy, LDLs are selectively adsorbed from the blood by an external adsorbent, and then the purified blood is re-introduced back into the patient's body \cite{Ronco2022}. Various types of the LDL adsorbents have been developed \cite{Sacks2003, Sikorski2007, Zhao2009, Gunkel2013, Yu2021, Dehghan2021a, Fang2024} with the main focus being their high selectivity towards LDLs, adsorbing effectiveness, scalability, and preferably renewability for a repeated usage.

Practical realizations of the selective LDL adsorbent include amphiphilic polymers \cite{Cheng2003, Yu2022} and biomimetic polymers \cite{Yu2021}, that are often grafted to a mobile carrier, e.g., magnetic nanoparticle \cite{Yu2022}. Renewability requires controlling of the adsorbance level by means of an external stimulus \cite{Ergun2025}. The examples of such systems, not necessarily related to the LDL, include: thermo-~\cite{Lanzalaco2017,Sudre2020,Yaremchuk2023}, magnetically-~\cite{Tian2023} and photo-controllable \cite{Malm2010} smart surfaces \cite{Bahl2020}. 

Recently, the photo-controllable LDL adsorbent has been developed which is characterised by high selectivity to LDL and reusability \cite{Guo2022, Guo2024}. Its photo-controllability is achieved by incorporation of the azobenzene chromophores \cite{Wei2015, DeMartino2020, Merritt2021} into a polymeric structure. Modification of phospholipid molecules by replacing one of its chains by  azobenzene chromophore or embedding azobenzene into a phospholipid membrane are known since at least 1980s--1990s \cite{Sandhu1986, Song1996}, and the interest to the photo-controllable membranes exploded recently \cite{Delova2025, Pritzl2025, Tomoshige2025, Guinart2025}. Under normal conditions (visible or no light), azobenzenes are in their non-polar \textit{trans} state, and mix well with the phospholipids from the external layer of the LDL, serving as binding sites \cite{Guo2022}. When illuminated by the UV light, azobenzenes photoisomerize into a polar \textit{cis} state causing unbinding LDLs and, hence, regeneration of an adsorbent. The initial tests of such adsorbent allowed the photo-regeneration up to 97.9\% of its initial capability after five adsorption/regeneration cycles \cite{Guo2022, Guo2024}.

This experimental setup motivated us to construct a closely related model that retains polymeric architecture of a photo-controllable adsorbent and simplifies the structure of the LDL particle down to a spherical core decorated by a layer of phospholipids \cite{Ilnytskyi2023}. The model is based on our previous studies of the azobenzene-containing polymers and decorated nanoparticles \cite{Ilnytskyi2006, Ilnytskyi2011, Ilnytskyi2015, Ilnytskyi2016p, Ilnytskyi2016m, Ilnytskyi2018, Ilnytskyi2019, Slyusarchuk2020, Yaremchuk2022}. Computer simulations, performed via molecular dynamics, allowed us to clarify the interrelation between the internal structure of a polymer brush and its binding strength. We found that, for each particular polymer length, there exists an interval of grafting densities characterized by a relatively high binding strength \cite{Ilnytskyi2023}. This regime is found in between the low grafting density case, characterized by insufficient number of azobenzenes to bind LDLs; and the high grafting density case, where LDLs cannot be bound because of the high density of a brush. Similar effects are found in some experimental works \cite{Xue2011}. 

The aim of the current study is to examine the possibility of a further improvement of the adsorption effectiveness (which includes both adsorption capacity and binding strength) by micropatterning the adsorbent structure. The effect of micropatterning of various kinds is discussed in a large number of applications, including adsorption of the LDLs \cite{Zhao2009, Dehghan2021b}, functionalized nanoparticles \cite{Steinbach2013}, proteins and cells \cite{Zhou2013, Wang2014, Kim2021, Aar2022, Badenhorst2024, Badenhorst2025, Mir2025}, as well as in the related computer simulations studies \cite{AdroherBentez2023, Ilnytskyi2025}. Micropatterned smart surfaces can be manufactured via a range of approaches \cite{Ma2019}, namely: atom transfer radical polymerization~\cite{Pyun2003, Edmondson2004, Ouchi2009, Ahmad2011, Steinbach2013, Zhou2013}, combined also with ring-opening polymerization~\cite{Johnson2010, Olivier2013}; light-excited controllable radical polymerization \cite{Zhao2022}; photo-lithography \cite{Welch2012, Yu2023, Badenhorst2024}; microcontact printing and surface-initiated polymerization \cite{Jeon1999}; block copolymer micelle lithography \cite{Wang2014, Badenhorst2025}, etc.

The aim of this study is to examine how the micropattern type and pitch affect the effectiveness of a photo-controllable polymer adsorbent, with the ultimate  aim to predict the setup with the highest adsorption effectiveness. To this end, we extend the model with uniformly grafted chains, suggested by us earlier \cite{Ilnytskyi2023}, for the case of bunch-like and grid-like micropatterns of variable pitch. These are examined for their adsorption effectiveness and compared against the adsorbent with uniformly grafted chains, at the same grafting densities of polymer chains. We establish the combination of the micropattern type, alongside with a set of related parameters for its structure, that leads to the highest effectiveness of the LDL adsorption. The outline of this study is as follows. In section~\ref{sec2} we describe the coarse-grained model for the micropatterned adsorbent and a LDL. Section~\ref{sec3} contains the results of computer simulations of this model and examination of the adsorption effectiveness depending on the micropattern type, pitch, grafting density, as well as the obtained adsorption isotherms. The study is finalised by conclusions.

\section{Modelling details}\label{sec2}

We first consider modelling of the LDL. These are known to be heterogeneous in size and density, and two general types, pattern A type, with the diameter $D^*_\mathrm{LDL}>25.5\,\mathrm{nm}$, and pattern B, $D^*_\mathrm{LDL}<25.5\,\mathrm{nm}$, are clearly distinguished \cite{Austin1986}. There are indications that the pattern B prevails in the blood of patients with coronary artery disease \cite{Campos1992, Austin1988, Stampfer1996, Lamarche1997}, possibly because of their prolonged residence in plasma and, consequently, a higher probability to be oxidized at the artery walls \cite{Scheffer1997}. This type of the LDL, with $D^*_\mathrm{LDL}\approx$ 20--23 $\mathrm{nm}$ \cite{Scheffer1997}, is the main target for selective adsorption. Experimentally, the spherical, discotic and ellipsoidal shapes are reported \cite{Segrest2001, VanAntwerpen1994, Teerlink2004, Ren2009}. One of the explanations for such discrepancy is the effect of the temperature, namely, at cryogenic conditions, essentially non-spherical shapes are detected, whereas at physiological temperatures, the LDLs appear more spherical in shape \cite{Kumar2011}, and we assume this shape in the current study.

\begin{figure}[htb] 
\begin{centering}
\includegraphics[height=4.5cm]{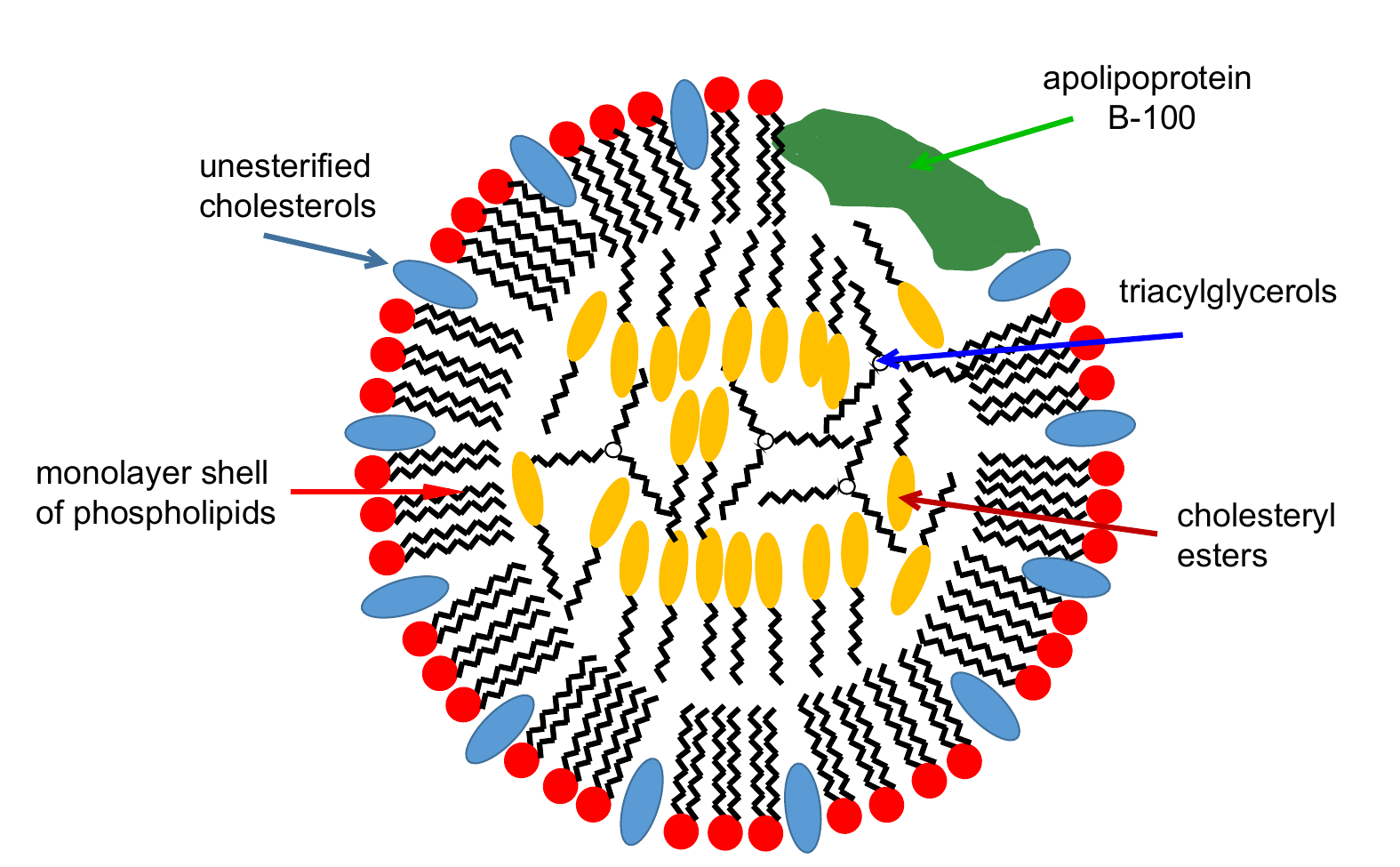}~~~
\includegraphics[height=3.95cm]{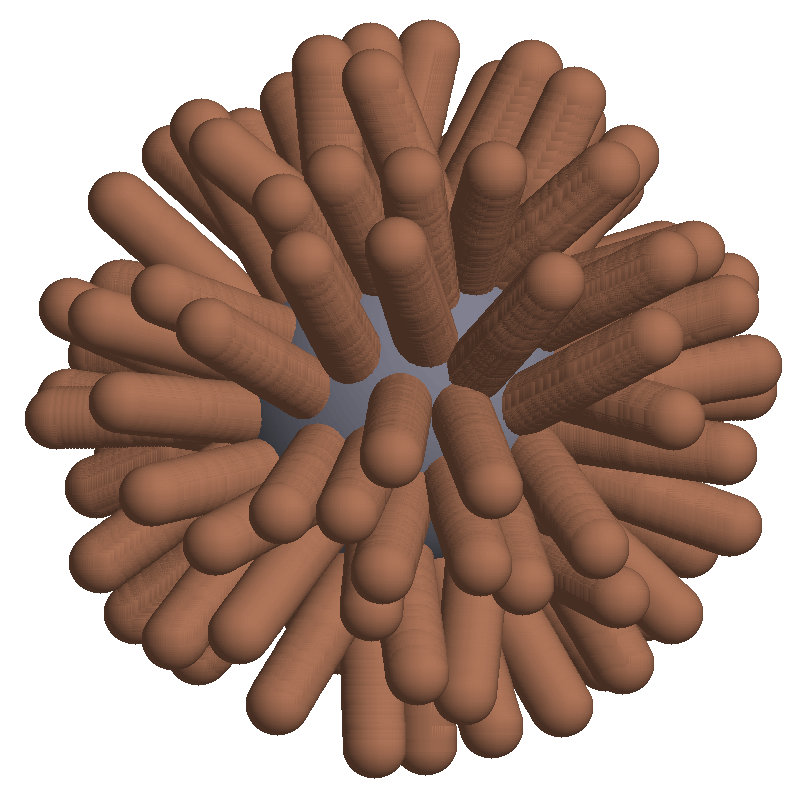}\\
\hspace{2cm}(a)\hspace{5.5cm}(b)\\
\end{centering}
\caption{(Colour online) (a) Schematic representation of the LDL composition, according to~\cite{Laguerre2007}. (b)~Coarse-grained model of LDL comprising a central sphere (representing LDL core of cholesterol esters and triacylglycerols) decorated by $100$ elongated particles (representing external monolayer shell of phospholipids). Unesterified cholesterols and apolipoprotein B-100 are not represented explicitly.}
\label{Model_LDL}
\end{figure} 

Following the paradigm of a coarse-graining approach, we retain only the fragments that are directly involved in the LDL adsorption. Based on the generally established view on the internal structure of the LDL \cite{Hevonoja2000, Laguerre2007, Prassl2008, Dai2023}, it consists of the core containing cholesterol esters and triacylglycerols and an external shell of phospholipids, cholesterols and Apolipoprotein B-100, see figure~\ref{Model_LDL} (a). At the pre-adsorption stage, the electrostatic interaction between the B-100 and specific groups of adsorbent play an important role in getting them in contact. However, after the contact between the two is established, the main anchoring effect came from the hydrophobic interactions between the phospholipids and hydrophobic groups of adsorbent, i.e., \textit{trans}-azobenzenes in the case of the experimental setup of~\cite{Guo2022}. The core of cholesterol esters and triacylglycerols defines the general shape of the LDL but does not participate in adsorption directly and, therefore, can be treated as a structureless sphere. In this way, the model LDL is reduced to the class of decorated nanoparticles, and we exploit our previous experience in this area \cite{Ilnytskyi2006, Ilnytskyi2011, Ilnytskyi2015, Ilnytskyi2016p, Ilnytskyi2016m, Ilnytskyi2018, Ilnytskyi2019, Slyusarchuk2020, Yaremchuk2022}. The diameter of a core sphere is set at $2.14\nm$. The outer shell of phospholipids, which participates in adsorption directly, is modelled as a collection of spherocylinder particles with the diameter of their spherical cap equal to $D=0.37\nm$ and the length-to-breadth ratio of $L/D=3$, yielding the spherocylinder length of $L\approx 1.5\nm$. Thus, the diameter of a model LDL is $D_\mathrm{LDL}\approx 2.14+2\cdot 1.5\nm=5.14\nm$, four times less than its real counterpart with the diameter  $D^*_\mathrm{LDL}\sim 20\nm$. This brings us to the relation between the model and the real life length scale of about $1\!:\!4$. Following the more detailed coarse-grained model of Murtola et al. \cite{Murtola2011}, surface packing fraction of spherocylinders is about $\eta\approx 0.79$, which in our model is achieved for the case of $100$ phospholipids per each LDL.
\begin{figure}[!h] 
	\begin{centering}
		\includegraphics[scale=0.4]{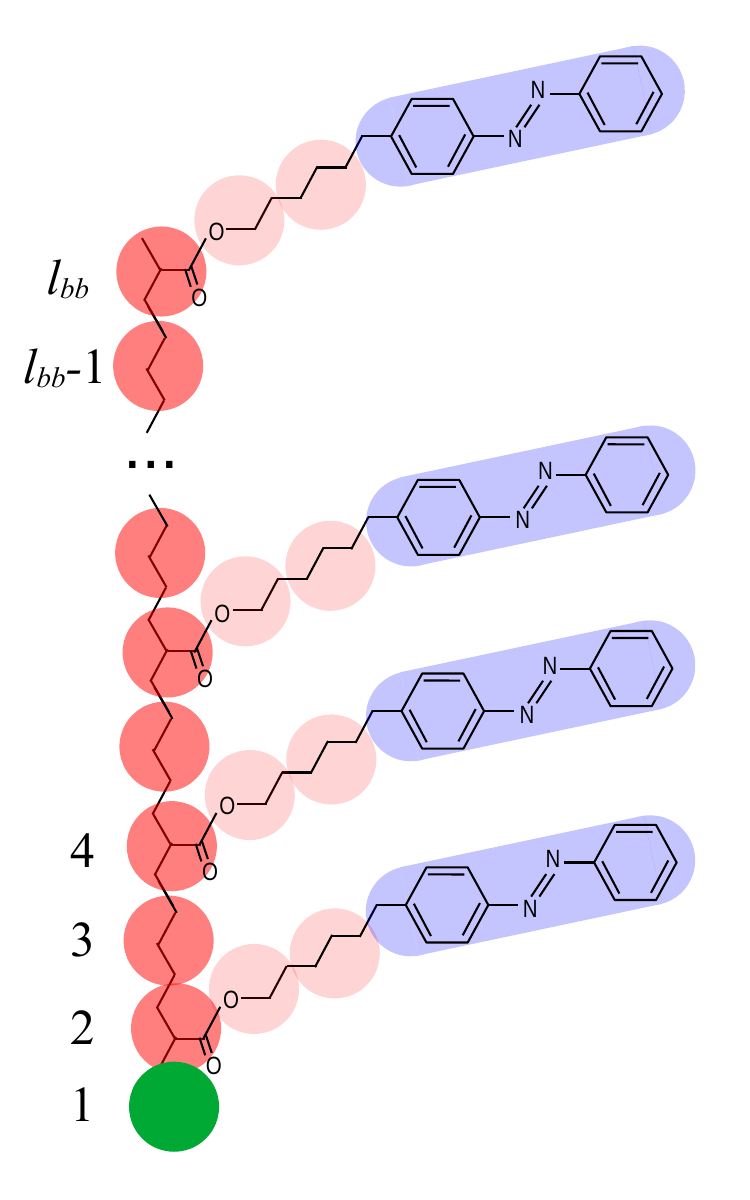}\\
	\end{centering}
	\caption{(Colour online) Coarse-graining of a single azobenzene-containing side-chain polymer. Each spherical bead of a backbone (shown in red), and of a side-chain (in pink) contains 3-4 heavy atoms, whereas an azobenzene chromophore is modelled via a single spherocylinder bead (shown in blue). The grafted group of a backbone is shown schematically as a green bead.}
	\label{Model_brush_chain}
\end{figure}

The photo-controllable adsorbent is modelled as a micro-patterned brush of azobenzene-containing polymers of the side-chain architecture \cite{Pan2008, Han2015, Liu2024, Kong2025}. Such polymers have been modelled on a coarse-grained level in a number of our previous studies \cite{Ilnytskyi2006, Ilnytskyi2011, Ilnytskyi2016m, Ilnytskyi2019}. In the current study, we consider polymers with backbone lengths of $l_{bb}=6$, $12$, and $24$ monomers represented by coarse-grained spherical beads with the diameter $\sigma=0.46\nm$, each mimicking a group of a few heavy atoms (carbon or oxygen) \cite{Hughes2005}. Each second monomer of a backbone serves as a branching point where a side chain, comprising a spacer of two monomers and a terminating azobenzene, is attached, see figure~\ref{Model_brush_chain}. Photocontrollability of the adsorbent is based on good miscibility of phospholipids with \textit{trans-}isomers of azobenzene, and their immiscibility with \textit{cis-}isomers, whereas the difference between the shapes of isomers play much lesser role. Therefore, we model both isomers by the same soft-core spherocylinder \cite{Ilnytskyi2015}. Another simplification, which is justified on a coarse-grained level, is to use the model spherocylinder for azobenzene of the same shape as the one used for a phospholipid. Azobenzenes are well known to integrate easily and effectively into lipid membranes for their photo-controllability \cite{Sandhu1986, Song1996, Delova2025, Pritzl2025, Tomoshige2025, Guinart2025}. The difference in the hydrophobicity of two azobenzene isomers is reflected in their respective interaction potentials with the phospholipids, as discussed below. Alternatively, one may use mechanically switched force fields \cite{Herrero2025}.

Each $i$\,th bead of any type is characterized by its center of mass position, $\vvec{r}_i$, whereas each spherocylinder bead has additionally its orientation, given by a unit vector $\vvec{e}_i$ collinear with its long axis. Then, in most general case, the mutual arrangement of two beads, $i,j$, is characterised by the set of variables $\vvec{q}_{ij}=\{\vhat{e}_i,\vhat{e}_j,\vvec{r}_{ij}\}$, where $\vvec{r}_{ij}=\vvec{r}_{i}-\vvec{r}_{j}$ is the vector connecting their centers. For spherical beads, the orientations are omitted. Following Kihara
\cite{Kihara1951,Kihara1963}, we treat all beads as convex bodies with an internal core, and evaluate the closest separation, $d(\vvec{q}_{ij})$, between the cores of two interacting beads.

The principal mechanism for binding LDLs to a polymer brush is attraction between their respective phospholipids and \textit{trans-}azobenzenes. This is the only attractive non-bonded interaction in our model, for which we employed the Lintuvuori and Wilson soft-core interaction potential \cite{Lintuvuori2008}, which can be written in the dimensionless form as
\begin{equation}\label{SAP}
V^{\mathrm{SAP}}_{\mathrm{NB}}[d'(\vvec{q}_{ij})]=\left\{
\begin{array}{ll}
U\big\{[1-d'(\vvec{q}_{ij})]^2-\epsilon'(\vvec{q}_{ij})\big\},& 0\leqslant d'(\vvec{q}_{ij})<1\vspace{2mm}\\
U\big\{[1-d'(\vvec{q}_{ij})]^2-\epsilon'(\vvec{q}_{ij})\big.&\\
\hspace{2em}\big.-\frac{1}{4\epsilon'(\vvec{q}_{ij})}[1-d'(\vvec{q}_{ij})]^4\big\},& 1 \leqslant d'(\vvec{q}_{ij}) \leqslant d'_c\vspace{2mm}\\
0,& d'(\vvec{q}_{ij})>d'_{c},
\end{array}
\right.
\end{equation}
where $U$ is the repulsion strength and defines the energy scale, $d(\vvec{q}_{ij})$ is the closest separation between the phospholipid and azobenzene cores, $d'(\vvec{q}_{ij})=d(\vvec{q}_{ij})/D$ is its dimensionless counterpart. The dimensionless, angle-dependent shift of this potential
\begin{equation}\label{eps}
\epsilon'(\vvec{q}_{ij})=\Big\{4\Big[U'_a-5\epsilon'_1 P_2(\vhat{e}_i\cdot\vhat{e}_j)-5\epsilon'_2\Big(P_2(\vhat{r}_{ij}\cdot\vhat{e}_i)+P_2(\vhat{r}_{ij}\cdot\vhat{e}_j)\Big)
\Big]\Big\}^{-1}
\end{equation} 
is obtained from the condition, that both the expression (\ref{SAP}) and its first derivative with respect to $d'(\vvec{q}_{ij})$ turn to zero when $d'(\vvec{q}_{ij})=d'_{c}$, where $d'_{c}=1+\sqrt{2\epsilon'(\vvec{q}_{ij})}$ is the cutoff separation for the potential \cite{Lintuvuori2008}. Here, $\vhat{r}_{ij}=\vvec{r}_{ij}/r_{ij}$, $U'_a$, $\epsilon'_1$, and $\epsilon'_2$ are dimensionless parameters that define its well depth, $\varepsilon$, and $P_2(x)=(3x^2-1)/2$ is the second Legendre polynomial. $\varepsilon$ is defined as $\epsilon'(\vvec{q}_{ij})$ for the side-to-side parallel arrangement between the interacting phospholipid-azobenzene pair
\begin{equation}\label{well-depth}
\varepsilon=U\left[4(U'_a-5\epsilon'_1+5\epsilon'_2)\right]^{-1}.
\end{equation}

All other pairs of beads interact via soft-core repulsive non-bonded potential
\begin{equation}\label{SRP}
V^{\mathrm{SRP}}_{\mathrm{NB}}[d'(\vvec{q}_{ij})]=\left\{
\begin{array}{ll}
U\left[1-d'(\vvec{q}_{ij})\right]^2, &0\leqslant d'(\vvec{q}_{ij})\leqslant 1\vspace{2mm},\\
0, &d'(\vvec{q}_{ij}) > 1,
\end{array}
\right.
\end{equation}
that is used typically in the dissipative particle dynamics simulations \cite{Groot1997}. The dimensionless closest separation between $i$\,th and $j$\,th beads is introduced as $d'(\vvec{q}_{ij})=d(\vvec{q}_{ij})/\sigma_{ij}$, where $\sigma_{ij}=(\sigma_i+\sigma_j)/2\;$ for two spheres of respective diameters $\sigma_i$ and $\sigma_j$, and $\sigma_{ij}=(\sigma_i+D)/2$ for the case when $j$\,th bead is a spherocylinder.

The total bonded interactions within the brush and within each LDL, respectively, are given as
\begin{equation}\label{Vbpol}
V^{\mathrm{BR}}_{\mathrm{B}}=\displaystyle\sum_{k=1}^{N_\mathrm{BR}}\left[\sum_{i=1}^{n'_b}k_b(l_i-l_0)^2 + \sum_{i=1}^{n'_a}k_a(\theta_{i}-\theta_0)^2 + \sum_{i=1}^{n'_z}k_z(\zeta_{i}-\zeta_0)^2\right],
\end{equation}
\begin{equation}\label{Vbldl}
V^{\mathrm{LDL}}_{\mathrm{B}}=\displaystyle\sum_{k=1}^{N_\mathrm{LDL}}\left[\sum_{i=1}^{n''_b}k_b(l_i-l_0)^2 + \sum_{i=1}^{n''_z}k_z(\zeta_{i}-\zeta_0)^2\right],
\end{equation}
where $N_\mathrm{BR}$ is the total number of polymer chains in a brush, and $N_\mathrm{LDL}$ is the number of LDL particles, $n'_b$, $n'_a$, and $n'_z$ are the numbers of bonds, branching angles, and terminal angles in a single polymer molecule, and $n''_b$ and $n''_z$ are the numbers of bonds and terminal angles in a single LDL. The bonds within LDL particles maintain an external phospholipid layer at a certain separation from the core center. The energy term involving the branching angles, $\theta_{i}$, in~(\ref{Vbpol}) ensures orthogonality of side chains to a local orientation of a backbone at the branching points. Similarly, correct angle, $\zeta_i$, between  the orientation of a spherocylinder and the bond by which it is connected to a spherical bead, is ensured in~(\ref{Vbpol}) and (\ref{Vbldl}) by the respective, $\zeta_i$-dependent, energy term \cite{Wilson1997}. 

\begin{figure}[htb] 
	\begin{centering}
		\includegraphics[height=4.5cm]{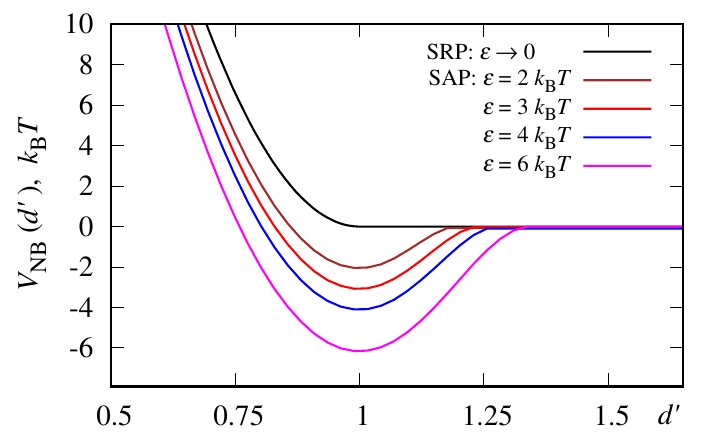}\\
	\end{centering}
	\caption{(Colour online) The shape of the repulsive potential, $V^{\mathrm{SRP}}_{\mathrm{NB}}$ (marked as ``SRP''), and that of the attractive potential, $V^{\mathrm{SAP}}_{\mathrm{NB}}$ (marked as ``SAP''), at various well depths, $\varepsilon$, in units of $k_{\text{B}}T$.}
	\label{Model_potential}
\end{figure}

The shape of both types of non-bonded interaction potentials, $V^{\mathrm{SAP}}_{\mathrm{NB}}$ and $V^{\mathrm{SRP}}_{\mathrm{NB}}$, are illustrated in figure~\ref{Model_potential}. The $V^{\mathrm{SAP}}_{\mathrm{NB}}$ potential is shown at several attraction strengths, given by the magnitude of its well depth, $\varepsilon$, see~(\ref{well-depth}), in units of $k_{\text{B}}T$.

%

\section{Results}\label{sec3}

We start from the choice for the attractive potential well depth, $\varepsilon$, see~(\ref{well-depth}) and figure~\ref{Model_potential}. In our previous work \cite{Ilnytskyi2023}, we used the set of parameters for the $V^{\mathrm{SAP}}_{\mathrm{NB}}$ potential representing the ``model A'' in~\cite{Lintuvuori2008}, the interaction potential also used in our previous studies \cite{Ilnytskyi2016p, Ilnytskyi2016m, Ilnytskyi2018, Ilnytskyi2019, Slyusarchuk2020, Yaremchuk2022}. However, it is characterized by a relatively deep well width of $\varepsilon=6k_{\text{B}}T$. The interaction as strong may turn adsorption of each LDL into an irreversible event, thus hampering equilibration of the system, especially for the case of a patterned brush.

To clarify this issue, we perform a series of simulations for the setup of~\cite{Ilnytskyi2023} with the well depths ranging from $\varepsilon=2\,k_{\text{B}}T$ to $6\,k_{\text{B}}T$. The simulation box of the dimensions $L_x=L_y=12.5\nm$ and $L_z=20\nm$ contained a total number of $N_\mathrm{ch}=108$ polymers uniformly grafted on both of $z=0$ and $z=L_z$ walls, yielding dimensionless polymer grafting density of $\rho_g=N_p\sigma^2/(2\,L_x\,L_y)\approx 0.73$ on each wall. The number of LDLs is $N_\mathrm{LDL}=18$, and these are initially arranged at the middle of the simulation box. Throughout the simulations, we focused on the time evolution of the fraction $f_\mathrm{bind}=N_\mathrm{bind}/N_\mathrm{LDL}$ of binded LDLs.

\begin{figure}[htb] 
	\begin{centering}
		\includegraphics[scale=0.7]{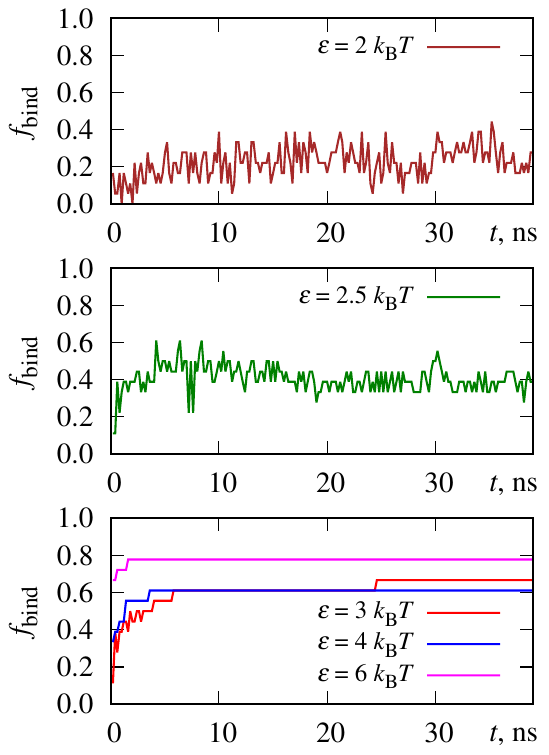}\\
	\end{centering}
	\caption{(Color online) Time evolution of the fraction $f_\mathrm{bind}=N_\mathrm{bind}/N_\mathrm{LDL}$ of bound LDLs at different well depth $\varepsilon$ of the phospholipid-azobenzene interaction. The system setup followed that of~\cite{Ilnytskyi2023}.}
	\label{fads_var_eps}
\end{figure}

The results are shown in figure~\ref{fads_var_eps}. At a shallow well depth, $\varepsilon\leqslant 2\,k_{\text{B}}T$, the fraction $f_\mathrm{bind}$ fluctuates strongly, indicating reversible binding of LDLs. However, the stationary state is characterized by a rather low value, $\bar{f}_\mathrm{bind} \sim 0.2$, in this case. On the contrary, at $\varepsilon\geqslant 3\,k_{\text{B}}T$, the stationary state for the fraction is high, from $0.6$ to $0.8$, but the time evolution is strictly monotonic, indicating irreversible binding of LDLs. The intermediate case, $\varepsilon=2.5\,k_{\text{B}}T$, seems to be a good compromise both displaying sufficiently high value of $\bar{f}_\mathrm{bind}\approx 0.4$, and the reversibility of the LDL binding.

\begin{figure}[htb] 
	\begin{centering}
		\includegraphics[height=2.5cm]{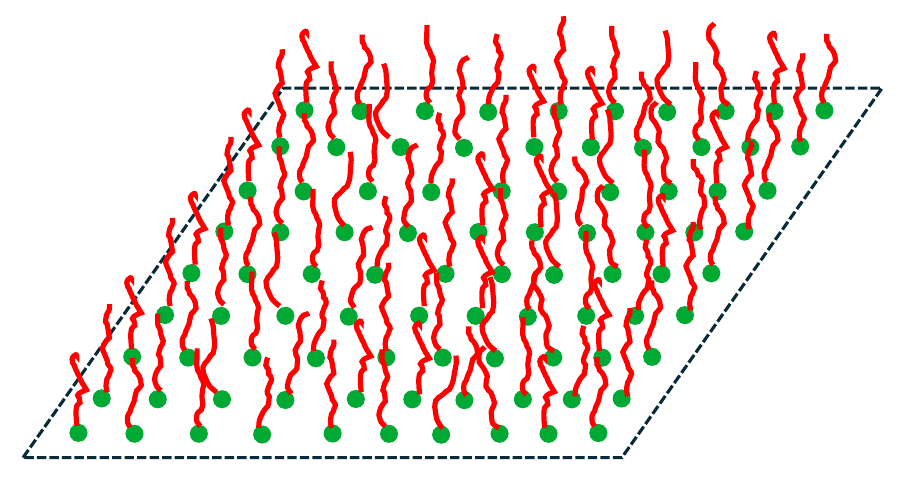}
		\includegraphics[height=2.5cm]{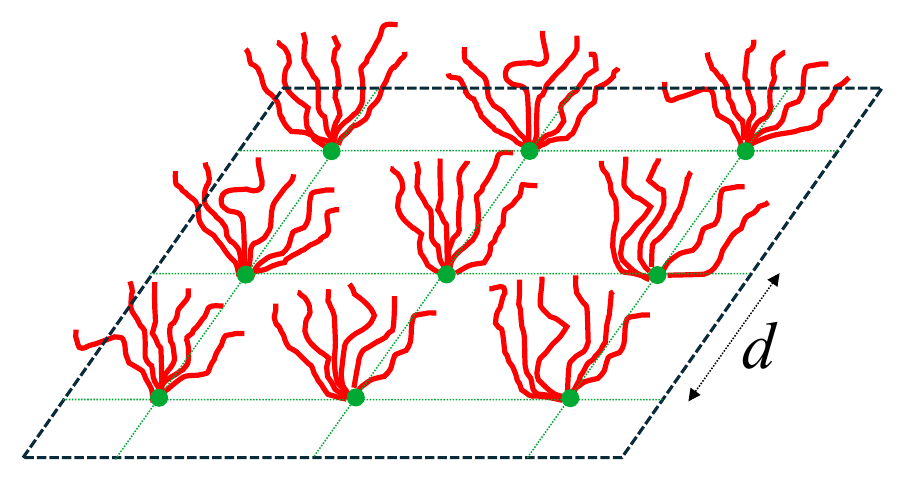}
		\includegraphics[height=2.5cm]{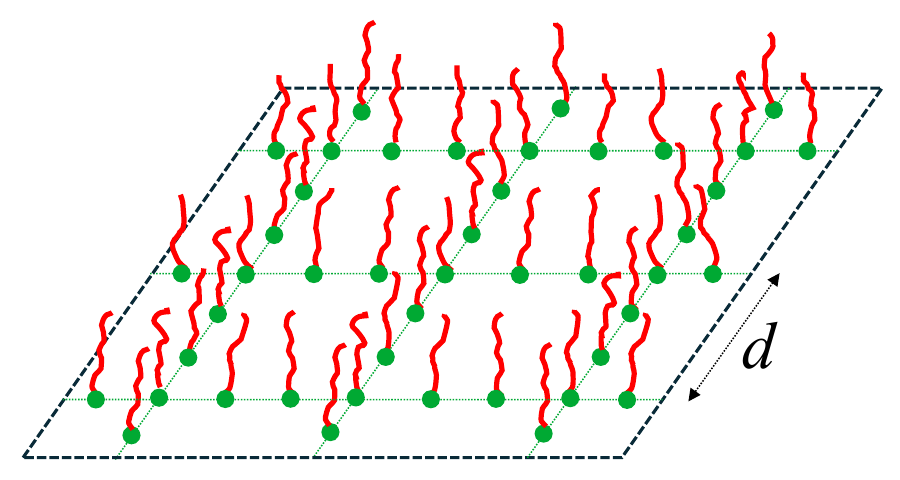}\\
		(a)\hspace{4.5cm}(b)\hspace{4.5cm}(c)~~~~~~~~~~\\
	\end{centering}
	\caption{(Colour online) Grafting micropatterns considered in this study. (a) Uniform brush case; (b)~bunch-like micropattern using regular grid points (shown as green discs); (c) grid-like micropattern with polymers evenly grafted along the lines connecting grid points. Characteristic pitch of the micropatterns (b) and (c) is denoted as $d$.}
	\label{Model_brush}
\end{figure}

As pointed out above, the main goal of this study is to examine the effect of spatial micropatterning of an adsorbent, and the set of micropatterns used here is illustrated in figure~\ref{Model_brush}. To avoid the picture clogging, each grafted side-chain polymer is shown by a wavy curve representing its backbone. The micropatterns examined include: an uniform brush, figure~\ref{Model_brush} (a), a bunch-like micropattern, figure~\ref{Model_brush} (b), and a grid-like micropattern, figure~\ref{Model_brush} (c).  The uniform brush, as the name suggests, contains $N_p=104$ polymers uniformly grafted to both walls. Both micropatterns, (b) and (c), are based on a square lattice of $3 \times 3$ sites with the lattice constant $d$, the lattice sites are shown in figure~\ref{Model_brush} as green discs. In the bunch-like micropattern, (b), a group of $6$ polymers are grafted to a single lattice site forming a ``bunch" that can also be interpreted as a single star-like polymer grafted by its central bead. The total number of grafted polymers is $N_p=108$. In the grid-like pattern, (c), polymers are grafted along the lines connecting lattice sites with their total number of $N_p=90$.

\subsection{Adsorption effectiveness depending on the polymer length, spatial pattern, and grafting density at constant LDL concentration}

The aim of this subsection is to examine the differences in adsorption effectiveness for the azobenzene-containing brush with various polymer lengths, grafting densities and micropatterns. To this end, we use the following protocol. The number of LDL particles, $N_\mathrm{LDL}$, is chosen first. Then, we evaluate the total area of a substrate, $A=2L_xL_y$ (of both bottom and top sides), such that the LDL particles could form a hexagonal lattice with maximum possible packing fraction of $f=\piup/\sqrt{12}$,
\begin{equation}\label{LxLy}
A\,f=N_\mathrm{LDL}\piup R_\mathrm{LDL}^2,
\end{equation}
where $R_\mathrm{LDL}=D_\mathrm{LDL}/2\approx 2.57\nm$. Expression (\ref{LxLy}) defines the simulation box dimensions $L_x$ and $L_y$ such that $2L_xL_y=A$. The box height, $L_z=20\nm$, is fixed. By this choice, the volume fraction of all LDLs, $f_\mathrm{LDL}=N_\mathrm{LDL}v_\mathrm{LDL}/V=2\,f\,v_\mathrm{LDL}(\piup R^2_\mathrm{LDL}L_z)^{-1}$ is the same in all simulation runs, since it is independent of $N_\mathrm{LDL}$. Here, $v_\mathrm{LDL}=20.18\nm^3$ is an estimated volume of a single model LDL, according to the dimensions of its constituents as given in section~\ref{sec2}. For each $N_\mathrm{LDL}$, the coordinates of grafting points are obtained by affine deformation of the respective micropattern shown in figure~\ref{Model_brush}, onto the walls with the total area of $A$. This means a distribution of the same total polymer mass over a smaller or larger substrate area. It is quantified by the polymer surface density
\begin{equation}\label{mgr}
\rho_{p} = \frac{N_{p}m_{p}}{A}=\frac{N_{a}(4m_1+m_2)}{A}=\rho_a~\alpha,
\end{equation}
where $N_p$ is the total number of  grafted side-chain polymers, $m_p$ is the mass of a single side-chain polymer, $N_a$ is the total number of azobenzenes in all grafted polymers. We also note that the number of spherical beads with a mass $m_1$ is four times larger than the number of azobenzenes with the mass $m_2$, see figure~\ref{Model_brush_chain}. Therefore, $N_{p}m_{p}=N_{a}(4m_1+m_2)$, we split $\rho_{p}$ in~(\ref{mgr}) into the dimensionless azobenzenes grafting density, $\rho_a$, and a model-dependent factor, $\alpha$
\begin{equation}\label{rho_a}
\rho_a=N_aD^2/A,~~~~\alpha = \frac{4m_1+m_2}{D^2},
\end{equation}
The adsorbent is traditionally characterized by its adsorption capacity
\begin{equation}\label{Eq:capac}
q = N_\mathrm{bind}\frac{m_\mathrm{LDL}}{N_{p}m_{p}},
\end{equation}
where $m_\mathrm{LDL}$ is the mass of a single LDL. Adsorption capacity per unit area, $q/A$, can be represented as
\begin{equation}\label{Eq:capac_red}
\frac{q}{A} = c~\beta,
\end{equation}
where we introduced the reduced adsorption capacity, $c$, and respective model dependent factor, $\beta$, as
\begin{equation}\label{Eq:capac_c}
c=1000\,f_\mathrm{bind}/N_a,~~~~~ \beta=\frac{f~m_\mathrm{LDL}}{1000~\piup R^2_\mathrm{LDL}(4m_1+m_2)}.
\end{equation}
A scaling factor of $1000$ is introduced to bring the working range of $c$ away from very small numbers.

Another important property of interest is the LDL binding strength, which is the sum of all binding energies, $V_{ij}<0$, between the phospholipid-azobenzene pairs $\langle i,j\rangle$ per single bound LDL,
\begin{equation}\label{Eq:Ebind}
  E_\mathrm{bind} = \frac{V_\mathrm{bind}}{N_\mathrm{bind}},\hspace{1em}
  V_\mathrm{bind} =\left| \sum_{\langle i,j \rangle}H\left(-V_{ij}\right)V_{ij}\right|,
\end{equation}
where $H(x)$ is the Heaviside step function, and the expression for the phospholipid-azobenzene interaction energy, $V_{ij}=V^{\mathrm{SAP}}_{\mathrm{NB}}$, is given by~(\ref{SAP}).

It also makes sense to combine the reduced adsorption capacity and the binding strength into a single characteristic, which can be termed as an adsorption effectiveness
\begin{equation}\label{Eq:eff}
  Q = c~ E_\mathrm{bind} = 1000\,\frac{f_\mathrm{bind}}{N_a}\frac{V_\mathrm{bind}}{N_\mathrm{bind}} = 1000\,\frac{V_\mathrm{bind}}{N_\mathrm{LDL}N_a}.
\end{equation}

\begin{figure}[htb] 
	\begin{centering}
		\includegraphics[height=3.8cm]{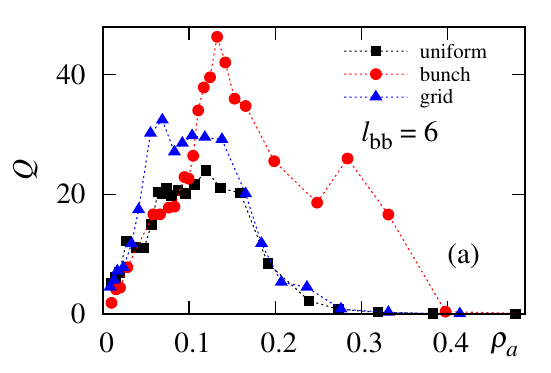}
		\includegraphics[height=3.8cm]{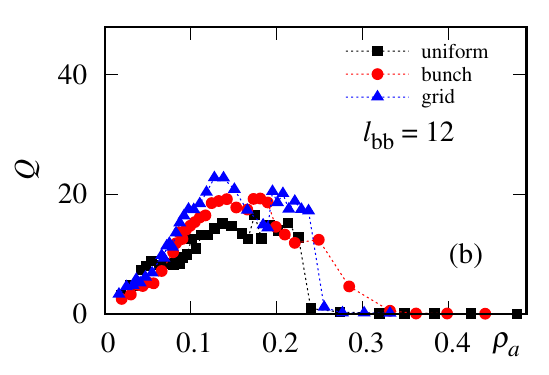}
		\includegraphics[height=3.8cm]{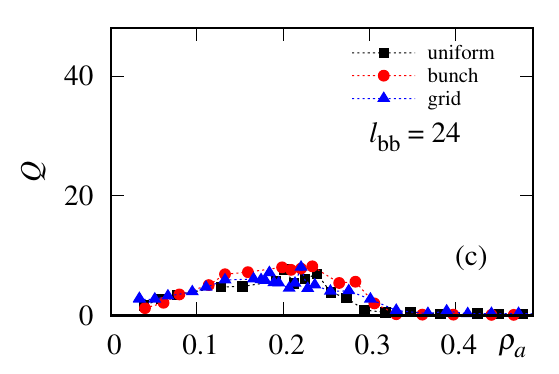}\\
	\end{centering}
	\caption{(Colour online) Adsorbent effectiveness, $Q$, see~(\ref{Eq:eff}), vs. dimensionless azobenzenes grafting density, $\rho_a$, see~(\ref{mgr}).  Three grafting patterns, introduced in figure~\ref{Model_brush}, are compared side-to-side in each plot. (a), (b), and (c) represent the cases of the backbone length of $l_{bb}=6$, $12$, and $24$, respectively.}
	\label{Fig:Q}
\end{figure}

It is plotted in figure~\ref{Fig:Q} as the function of grafting density, $\rho_a$. Three backbone lengths, $l_{bb}=6$, $12$, and $24$, are examined, and all three patterns, introduced in figure~\ref{Model_brush}, are analysed in each case. Here, we note two principal features. The first one is that an essential difference between three grafting patterns is observed for the case of the shortest backbone, $l_{bb}=6$, only, see figure~\ref{Fig:Q} (a). In this case, the values of $Q$ for the bunch-like pattern are almost always higher than these for the two remaining patterns, and display two distinct maxima. At the low density maximum, at $\rho_a\approx 0.13$, the bunch-like pattern is twice more effective than a uniform one, and $1.5$ times than the grid-like pattern. The high density maximum, $\rho_a\approx 0.28$, is distinct for the bunch-like pattern, and is absent in two other cases. The values for $Q$ for the grid-like and uniform patterns overlap at larger densities, $\rho_a>0.16$, whereas around the density of $\rho_a\approx 0.1$, grid-like pattern is about $1.4$ times more effective. This feature indicates an essential role played by the spatial structuring of a brush in the effectiveness of the LDL binding for the short backbones case. This effect is reduced for the case of longer backbone, $l_\mathrm{bb}=12$, see figure~\ref{Fig:Q} (b), and completely disappears for $l_\mathrm{bb}=24$, see figure~\ref{Fig:Q} (c). This can be explained by decorrelation between the structure of the top surface of a brush, which serves as the adsorbing surface for LDL, and the grafting micropattern on a substrate for sufficiently long chains.

The second feature is an essential decrease of the adsorption effectiveness, $Q$, with the increase of the backbone length from $l_\mathrm{bb}=6$ to $12$ and then further to $24$. This effect can be explained by the fact that only the utmost layer of a brush adsorbs LDLs. Therefore, the total binding energy, $V_\mathrm{bind}$, stays approximately the same with increase of  $l_\mathrm{bb}$, whereas $N_a$ grows up, thus decreasing $Q$, according to~(\ref{Eq:eff}).

\subsection{Analysis of the spatial arrangement of adsorbed LDLs}

\begin{figure}[htb] 
	\begin{centering}
		\includegraphics[height=2cm]{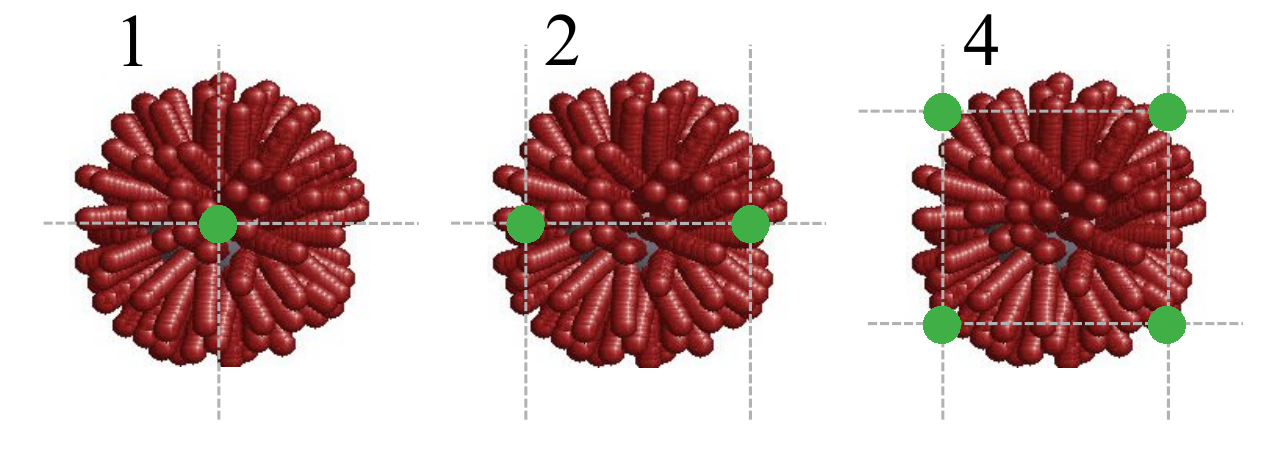}\\
		(a)\\
		~~~~~~\includegraphics[height=5cm]{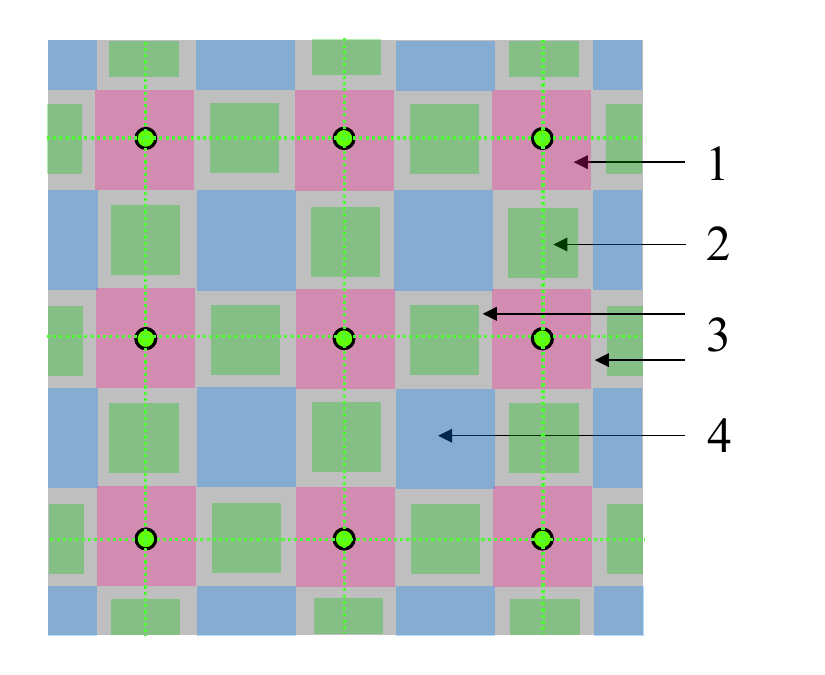}\\
		(b)\\
	\end{centering}
	\caption{(Colour online) (a) Definition of the LDL arrangement types $1$, $2$, and $4$. (b) Split of a substrate with grafted polymers into colored regions of equal total area to sort all LDLs according to their arrangement type. Grafted polymer beads are shown via small green circles.}
	\label{arr_def_sort}
\end{figure}

Because of the highest effectiveness of the adsorbent comprising short chains, $l_\mathrm{bb}=6$, and the strongest influence of the grafting micropatterning for such a brush, we consider this case more in  detail. Given the non-monotonous behaviour of $Q$ in the case of the bunch-like pattern, see figure~\ref{Fig:Q} (a), one expects that this pattern provides means for some energetically favorable arrangements of LDLs relatively to the grafted bunches of chains. To verify this scenario, we split all possible relative positions of LDLs into four types. Relative positions for LDL, classified as the arrangements of the types $1$, $2$, and $4$, are shown in figure~\ref{arr_def_sort} (a); all the rest LDL positions belong to the type $3$ arrangement. Sorting of LDLs is performed by splitting the substrate into colored regions built upon the lattice sites with grafted bunches, as shown in figure~\ref{arr_def_sort} (b). Respective areas for all types of arrangements are designated in the plot. The split is built in such a way that the areas occupied by each arrangement type are the same.

\begin{figure}[!thb] 
	\begin{centering}
		\hspace{1em}\textit{bunch-like pattern}\hspace{8em}\textit{grid-like pattern}\\
		\includegraphics[height=4.5cm]{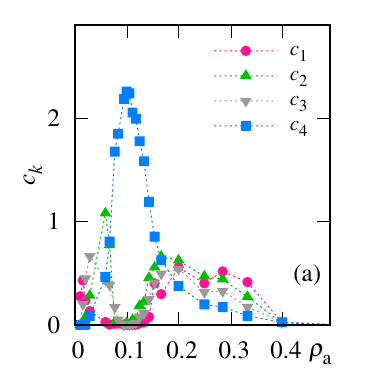}\hspace{2em}
		\includegraphics[height=4.5cm]{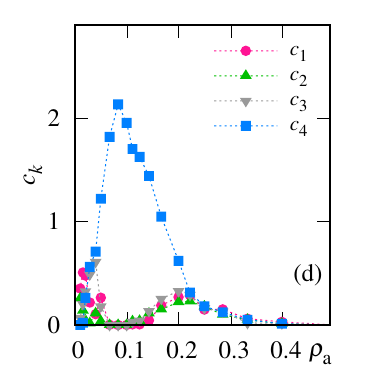}\\
		\includegraphics[height=4.5cm]{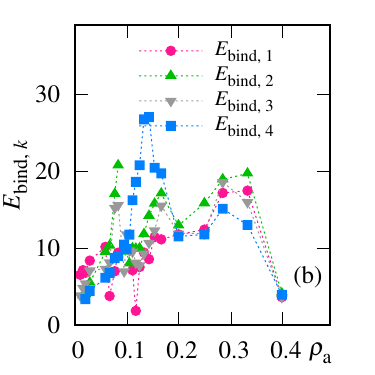}\hspace{2em}
		\includegraphics[height=4.5cm]{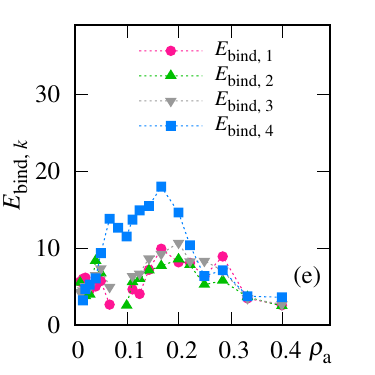}\\
		\includegraphics[height=4.5cm]{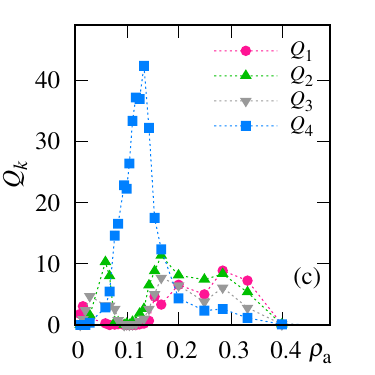}\hspace{2em}
		\includegraphics[height=4.5cm]{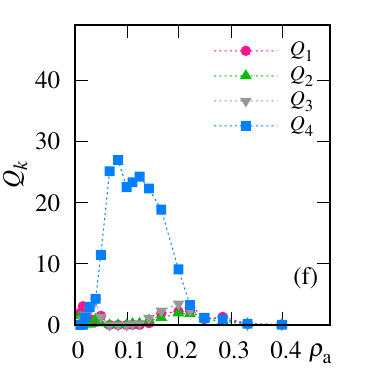}\\
	\end{centering}
	\caption{(Colour online) Partial reduced adsorbing capacities, $c_k$, binding strengths, $E_{\mathrm{bind}, k}$, and adsorbent effectivenesses, $Q_k$, for each $k$th type of arrangement of LDLs, see figure~\ref{arr_def_sort}, the case of $l_{bb}=6$. (a), (b), and (c) are for the case of bunch-like pattern; (d), (e), and (f) are for the case of grid-like pattern.}
	\label{arr_distr}
\end{figure}

Such a split enables us to monitor partial reduced adsorbing capacities, $c_k$, partial binding strengths, $E_{\mathrm{bind}, k}$, and partial adsorbent effectivenesses, $Q_k$, of each $k$th type of arrangement. Such plots are shown in figure~\ref{arr_distr}, where we compare these characteristics obtained for the case of a bunch-like pattern, plots (a), (b), and (c), with their counterparts for the grid-like pattern, plots (d), (e), and (f). The case of an uniform pattern is trivial and produces the same curves for all $k$th arrangement types, within statistical accuracy and is not shown. Comparison between bunch-like and grid-like patterns reveals much similarity, in particular, the prevailance of partial contributions from the arrangement type $4$ at low values of $\rho_a\leqslant 0.2$. For higher azobenzenes grafting densities, $\rho_a > 0.25$, the partial contributions from all arrangement types are comparable in their magnitude. The shapes of the curves for $c_4$, $E_{\mathrm{bind}, 4}$, and $Q_4$ at $\rho_a\leqslant 0.2$ for these two patterns differ, and the ones for the bunch-like pattern are narrower. The maxima, achieved by $E_{\mathrm{bind}, 4}$ and $Q_4$, are also visibly higher for the case of bunch-like pattern. This effect can be explained by the presence of the areas with very high local grafting density for the case of bunch-like pattern. Then, a narrow interval, $0.09<\rho_a<0.15$, for the characteristic pattern pitch, $d=(D/3)\sqrt{N_a/(2\rho_a)}$, exists where both $c_4$ and $E_{\mathrm{bind}, 4}$ display sharp narrow maxima, see figure~\ref{arr_distr} (a) and (b). This results in sharp narrow maxima for $Q_4$, see figure~\ref{arr_distr} (c). The sharp maximum for $Q_4$ at about $\rho_a<0.15$, is observed at the pattern pitch of $d\approx 4.1\nm=0.8\,D_\mathrm{LDL}$. For the case of grid-like pattern, distribution of grafting points is much more uniform, resulting in the spread of the bell-like shapes for both $c_4$ and $E_{\mathrm{bind}, 4}$ over a wider interval of $0.02<\rho_a<0.2$. The resulting shape of the $Q_4$ curve in this case has a flat top region spanning from $\rho_a=0.05$ to $0.15$. These densities are translated into the pattern pitches from $d=3.75\nm=0.73\,D_\mathrm{LDL}$ to $6.5\nm=1.26\,D_\mathrm{LDL}$. Therefore, this pattern is found to be effective in a wider interval of grafting densities, but exhibits a lower maximum effectivity compared to the case of bunch-like pattern.

We note that both micropatterns, considered here, display quite similar heights of the maxima for $c_4$, see figure~\ref{arr_distr} (a) and (d). Higher values for the adsorption effectiveness, $Q_4$, for the bunch-like pattern, see plots (c) and (f) of the same figure, are attributed then to the higher maxima heights for $E_{\mathrm{bind}, 4}$, see figure~\ref{arr_distr} (b) and (e). One should also note a higher contribution from $E_{\mathrm{bind}, 2}$ for the case of a bunch-like pattern. Hence, at the optimum grafting densities, bunch-like pattern provides approximately the same adsorbing capacity as grid-like pattern, but stronger binding of LDLs. The predominant contribution from the LDLs arrangement $4$ indicates that LDLs are ``nesting'' in between four bunches. One may suggest that a careful choice for a bunch-like micropattern pitch, $d$, may provide the mechanism for the dimension-based separation of the LDLs from the high density lipoproteins. Exact positions for grafting points could be difficult or costly to control experimentally, but one may be able to choose a particular grafting density which is characterized by the average separation between the bunches close to the required value of $d$.

The origin of the high density peak for the adsorption efficiency, $Q$, observed at $\rho_a\approx 0.28$ for a bunch-like micropattern 
in figure~\ref{Fig:Q} (a), cannot be attributed to a prevalence of any particular LDLs arrangement type. As follows from figure~\ref{arr_distr}, it is observed mainly in the behavior of $E_{\mathrm{bind}, k}$, see frame~(b). All four types of arrangement are found to contribute approximately equally, and their contributions are summed up. The grid-like micropattern does not have this feature.

The photoregeneration of the adsorbent occurs when the latter is illuminated by an ultraviolet light~\cite{Guo2022}. In this case, the azobenzenes photo-isomerize into the polar \textit{cis}-state, and this isomer does not attract phospholipids. In the realms of our model, this means that the phospholipid-azobenzene attractive potential,~(\ref{SAP}), is replaced by the soft repulsive potential,~(\ref{SRP}). As a result, LDLs are gradually desorbing until free in bulk. This process is independent of the adsorbent micropatterned and is described in detail in our previous work \cite{Ilnytskyi2023}, and, therefore, is not repeated here.

\subsection{Adsorption isotherms}

Previous sections cover computer simulations of the azobenzene-containing adsorbent at various surface densities, but at fixed concentration of LDLs. We found that the bunch-like grafting pattern of short polymers with $l_\mathrm{bb}=6$ at the azobenzenes grafting density around $\rho_a= 0.12$ is characterized by the highest adsorption effectiveness, $Q$, see figure~\ref{Fig:Q}. The reason for this is the prevalence of specific arrangement of LDLs with respect to the micropattern, see figures~\ref{arr_def_sort} and \ref{arr_distr}. In this section we fix both $l_\mathrm{bb}$ and $\rho_a$ for the adsorbent, and cover a wide range for the LDLs concentrations. This allows to compare the adsorption isotherms between the bunch-like and uniform grafting patterns. The simulation box dimensions are fixed at $L_x=L_y=13.5\nm$ and $L_z=20\nm$, the backbone length is $l_\mathrm{bb}=6$, and the resulting azobenzenes grafting density is $\rho_a= 0.12$. The number of LDLs is varied from $N_\mathrm{LDL}=2$ to $32$, resulting in the volume fractions of $0<f_\mathrm{LDL}<0.18$.
\begin{figure}[htb]
	\begin{centering}
		\includegraphics[height=4cm]{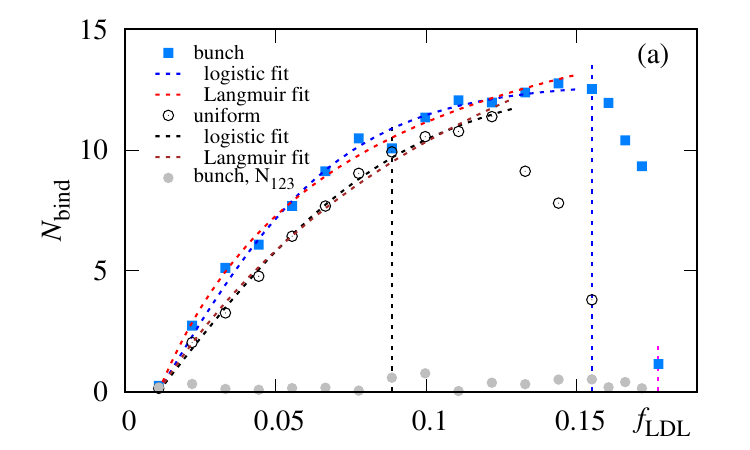}
		\includegraphics[height=4cm]{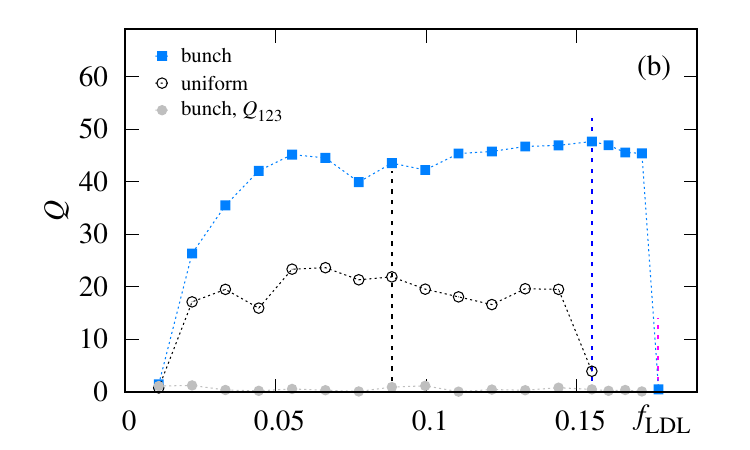}\\
		\includegraphics[height=4.5cm]{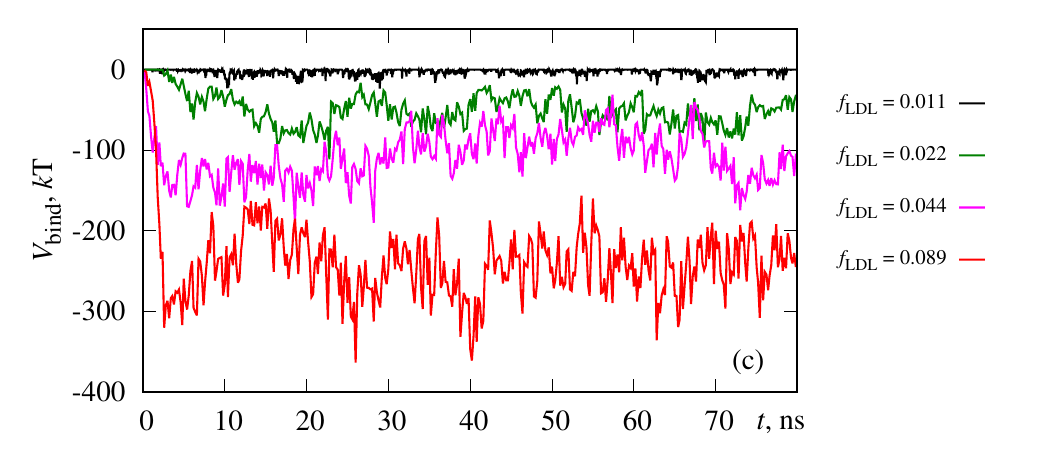}\\
	\end{centering}
	\caption{\label{Nb_Q_ads_isot}(a) Adsorption isotherms, in terms of $N_\mathrm{bind}$ as the function of $f_\mathrm{LDL}$, for both bunch-like and uniform patterns (as indicated in the plot), $N_{123}$ is explained in the text. Fits of the data to the Langmuir and logistic growth forms are shown via dashed color curves (as indicated in the plot). (b) The same for the adsorption effectivity, $Q$, but no fitting is provided in this case. Vertical dashed lines: black line at $f_\mathrm{LDL}\approx 0.09$ is the fixed LDL volume fraction in sections A and B; blue and magenta lines, at $f_\mathrm{LDL}\approx 0.155$ and $0.177$, respectively, mark the end of the ascending and descending regimes, respectively. (c) Fluctuations of the total binding energy, $V_\mathrm{bind}$, along the time trajectory, at various $f_\mathrm{LDL}$.}
\end{figure}

Since the molecular weight of a polymeric adsorbent is the same for all considered $f_\mathrm{LDL}$, adsorption capacity $q$,~(\ref{Eq:capac}), changes only with the number of bound LDLs, $N_\mathrm{bind}$. In this case, adsorption isotherms can be expressed as the functional dependencies of $N_\mathrm{bind}$ on $f_\mathrm{LDL}$. For the case of the bunch-like pattern, we also evaluate $N_{123}=\sum_{k=1}^{3}N_{\mathrm{bind},k}$, the contribution to $N_\mathrm{bind}$ from the spatial arrangements of LDLs of the types $1$, $2$, and $3$, see figure~\ref{arr_def_sort}. The datapoints for $N_\mathrm{bind}$ for the case of the bunch-like and uniform patterns are compared in figure~\ref{Nb_Q_ads_isot} (a). There are two regimes in both cases: (i) ascending regime, and (ii)~descending regime. In ascending regime, there is no LDLs crowdiness, and $N_\mathrm{bind}$ can be fitted reasonably well by the widely used Langmuir form, which is valid for a single adsorption layer and an uniform adsorbing surface
\begin{equation}\label{Eq:Langm}
N_{\mathrm{bind}} \approx \frac{N_\mathrm{max}\,K\,(f_\mathrm{LDL}-\Delta)}{1+K\,(f_\mathrm{LDL}-\Delta)}.
\end{equation}
Here $N_\mathrm{max}$ is some extrapolated maximum, $K$ is the Langmuir adsorption constant, and {$\Delta=0.011$ is the threshold volume fraction at which adsorption is started to be observed. Its non-zero value must be attributed to a highly dynamic binding/unbinding cycles, as evidenced by the fluctuations of total binding energy $V_\mathrm{bind}$, see figure~\ref{Nb_Q_ads_isot} (c). Unbinding must be dominating at a very low volume fraction of LDLs because binding is a rare event in this case.} Fitting data to this form yields $K=15.8$ for the bunch-like, and $K=7.45$ for the uniform grafting patterns. We do not observe some firm indication of structural changes in the adsorption isotherm, except for some local ``zig-zag'' at around $f_\mathrm{LDL}\approx 0.09$.

As far as the adsorbing area of a brush is finite, the number of bound LDLs, $N_{\mathrm{bind}}$, eventually saturates. This reminds the population growth models, \cite{Frauenthal1979, Price2023} where individuals compete for the live resources. It is logical, therefore, to assume that the time evolution for $N_{\mathrm{bind}}$ may follow the solution for this type of models, namely, the logistic growth function, which can be written in our case as
\begin{equation}\label{Eq:logistic}
N_{\mathrm{bind}} \approx N'_\mathrm{max} \frac{1-\re^{-K'(x-\Delta)}}{1+\re^{-K'(x-\Delta)}}.
\end{equation}
As a result of the fit, the following values for the logistic growth rate, $K'$, are found:  $K'=32.5$ for the bunch-like, and $K'=24.4$ for the uniform grafting patterns. For both types of fits, equations~(\ref{Eq:Langm}) and (\ref{Eq:logistic}), the respective constants $K$ and $K'$ for the bunch-like pattern are essentially higher than their counterparts for the uniform pattern. The results for both types of fits are shown in figure~\ref{Nb_Q_ads_isot} (a) via dashed lines and are marked accordingly. We also note that, for the bunch-like micropattern, the contribution $N_{123}$ from the arrangement types $1$, $2$, and $3$ to $N_\mathrm{bind}$ is negligibly small. Hence, almost all contributions to $N_\mathrm{bind}$ are provided by the spatial arrangement of LDLs of the type $4$.

We also show a similar plot for the adsorption effectiveness, $Q$, see figure~\ref{Nb_Q_ads_isot} (b). Contrary to the behavior of $N_\mathrm{bind}$, the dependence of $Q$ is found to be rather flat in a whole ascending regime. Typical values for $Q$ for the bunch-like pattern are about twice higher than these for the uniform pattern, throughout all this interval. A black dashed vertical line in both frames of figure~\ref{Nb_Q_ads_isot} indicates the LDL volume fraction, $f_\mathrm{LDL}\approx 0.09$, for which the simulations are performed in previous sections, A and B. One may note that the local ``zig-zag'' occurs at exactly this volume fraction as well, but most likely this is a mere coincidence. The other two vertical dashed lines, blue and magenta, indicate the end of the ascending and descending regimes, respectively, for the bunch-like micropattern.

In the descending regime, which is observed at sufficiently high $f_\mathrm{LDL}$, $N_\mathrm{bind}$ decays monotonously. For the uniform grafting pattern, this occurs at about $f_\mathrm{LDL}>0.125$, whereas for the grid-like grafting density: at $f_\mathrm{LDL}>0.155$. In contrast to this behavior, adsorption effectiveness, $Q$, drops to zero rather abruptly at $f_\mathrm{LDL}\approx 0.145$ and $f_\mathrm{LDL}\approx 0.17$, respectively.

\begin{figure}[htb] 
	\begin{centering}
		\includegraphics[height=4.5cm]{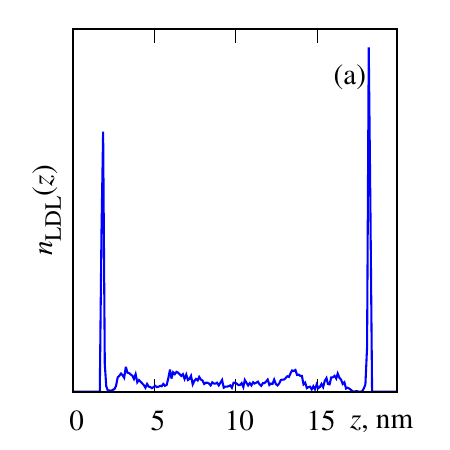}\hspace{1.5em}
		\includegraphics[height=4.5cm]{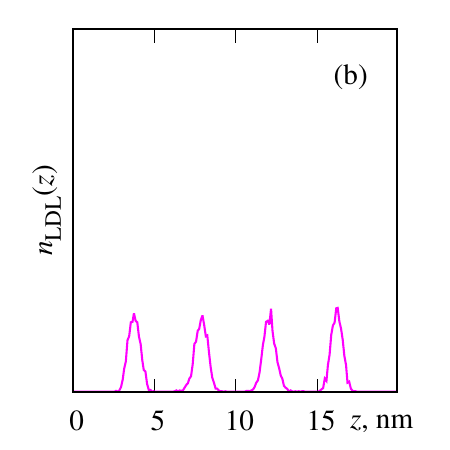}\\
		\includegraphics[height=4.5cm]{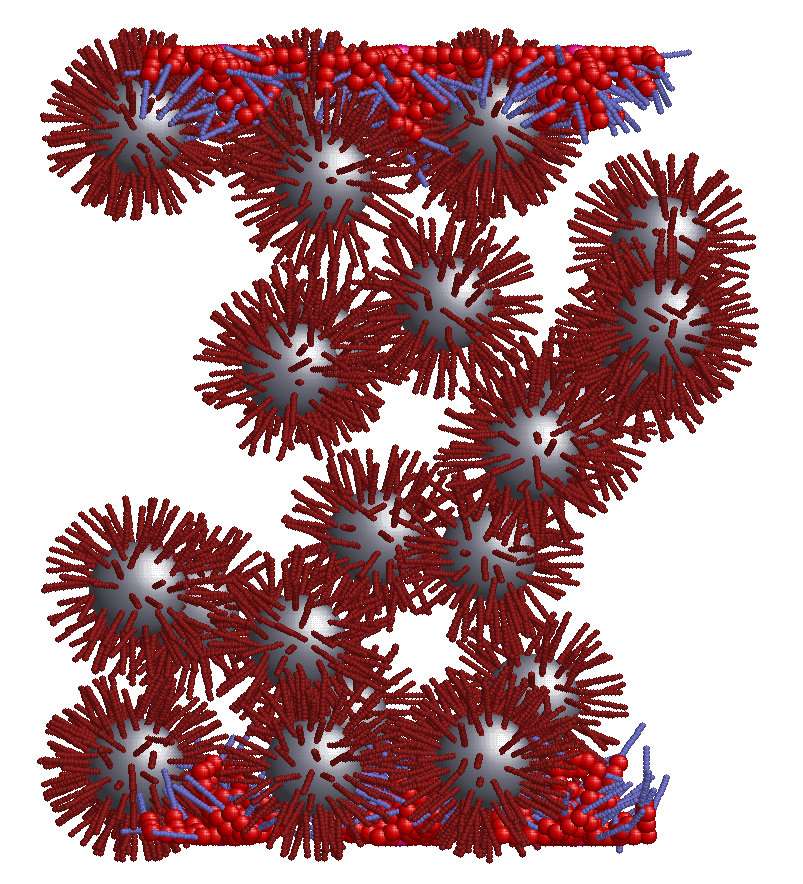}\hspace{2.5em}
		\includegraphics[height=4.5cm]{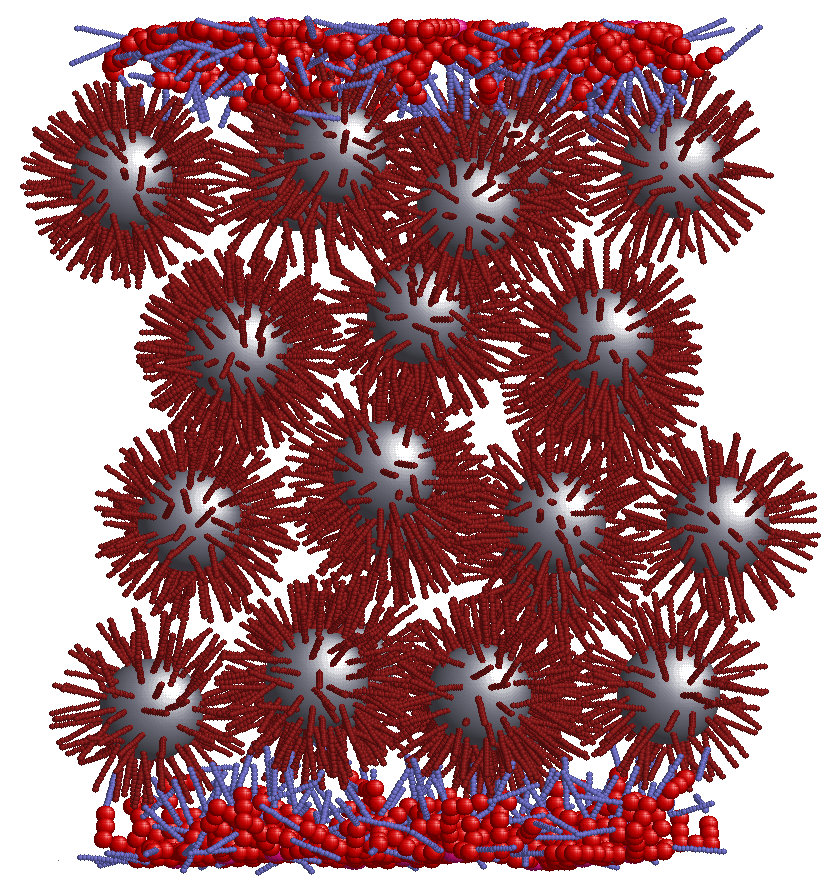}\\
		(c)\hspace{13em}(d)\\
	\end{centering}
	\caption{(Colour online) (a) Density profile, $n_\mathrm{LDL}(z)$, for the LDLs centers of mass along the $OZ$ axis at the LDLs volume fraction of $f_\mathrm{LDL}=0.155$ (the end of an ascending regime for a bunch-like micropattern). (b) the same at $f_\mathrm{LDL}=0.177$ (the end of a descending regime for the same micropattern). (c) Snapshot of the LDLs arrangement for the case (a), (d) the same for the case (b).}
	\label{snaps_ads_isot}
\end{figure}

To examine the details of the LDLs arrangement, we built the density profiles, $n_\mathrm{LDL}(z)$, for the LDLs centers of mass along the $OZ$ axis. The profiles are averaged over the last 40~ns of the simulations. The case of a bunch-like micropattern is considered at characteristic volume fractions, $f_\mathrm{LDL}=0.155$ and $0.177$, each marks the end for an ascending and descending regime, respectively. These are shown in figure~\ref{snaps_ads_isot} (a) and (b). Two sharp peaks in figure~\ref{snaps_ads_isot} (a) reflect two dense immobilized layers of bound LDLs, whereas $n_\mathrm{LDL}(z)$ is uniformly smeared for all $z$ in between them, reflecting high diffusivity of LDLs in a bulk during the time of observation. Figure~\ref{snaps_ads_isot} (b) shows four equally separated maxima of the same height indicating the formation of four layers. These findings are supported by two representative snapshots for both cases, as shown in figure~\ref{snaps_ads_isot} (c) and (d). While the arrangement of LDLs in a bulk in the frame (c) is rather loose, the frame (d) shows a highly ordered bulk cubic phase. It competes with the LDLs adsorption, and should be the main reason for its disappearance in this volume fraction of LDLs.

The situation is very similar for the case of an uniform micropattern (not shown). The crossover between the ascending and descending regimes occurs at a lower fraction of $f_\mathrm{LDL}\approx 0.125$, whereas the descending regime terminates at about $0.155$. These values are about $1.2$ times lower than their counterparts for the case of a bunch-like micropattern. This can be explained by the larger volume accessible to LDLs in the latter case because of the presence of ``nests'' (arrangement type $4$). These are absent for the case of an uniform pattern, in which case the brush thickness is non-zero and uniform throughout the adsorbing area. Our estimates made from the density profiles, $n_\mathrm{LDL}(z)$, built for both micropatterns support this simple explanation.

\section*{Conclusions}

Reducing the risk for cardiovascular diseases is one of the top priorities in medicine, affecting not only the health well-being, but the very sustainability of human lives. One of the factors that increases such a risk is the excess concentration of the LDLs in blood plasma. Hemoperfusion therapy, based on selective adsorption of LDLs outside the patient's body and its reintroduction back into veins, provides a strong alternative to the pharmacy-based medication using statins, ezetimibe, or fibrates, especially when such usage leads to strong side-effects. The main requirements to the adsorbent, involved in hemoperfusion, are as follows: the selectivity to LDL, effectiveness, low cost, and, preferably, reusability. The photo-controllable LDL adsorbent has been recently, developed which possesses these properties. It comprises spherical particles decorated by the azobenzene-containing polymer brush. Under normal conditions (visible or no light), the azobenzenes are in their hydrophobic \textit{trans} isomeric state, and form physical bonds with phospholipids from the external layer of LDLs due to a strong van der Waals hydrophobic interaction. Under ultraviolet light, azobenzenes photo-isomerize into a highly polar \textit{cis} state, resulting in breakage of these bonds and releasing LDLs free. In this way, the adsorbent is cleaned and can be reused repeatedly. Experimental studies report the photo-regeneration up to 97.9\% of its initial capability after five adsorption/regeneration cycles \cite{Guo2022, Guo2024}.

In our previous study \cite{Ilnytskyi2023}, we developed simplified mesoscopic models for both LDL and the azobenzene-containing polymer brush. Only the atomic groups that are directly involved in adsorption (phospholipids, azobenzenes, and the monomers of polymeric molecules) are described explicitly, whereas the LDL core is replaced by a structureless sphere. The uniformly grafted brush case was considered only, and the model length scale is reduced fourfold compared to real dimensions. Molecular dynamics simulations at constant LDL concentration and variable polymer lengths and grafting densities reveal that, for each particular polymer length, there exists an interval of optimal grafting densities characterized by a relatively high magnitude of the binding energy of LDLs \cite{Ilnytskyi2023}, in agreement with some experimental works \cite{Xue2011}. The current study extends both the adsorbent setup and the analysis of the adsorption process in a number of ways. 

First of all, we focus on the strength of the phospholipid-azobenzene attraction, which drives LDL binding. The optimal choice for the well depth of this interaction is found to be about $\varepsilon = 2.5k_{\text{B}}T$, in which case the binding of LDLs is reversible, allowing to reach a non-zero stationary concentration of the bound LDLs. 

We consider three distinct grafting patterns of azobenzene polymers in a brush, namely: bunch-like, grid-like, and an uniform pattern. These can be manufactured via a range of processes \cite{Ma2019} and have a profound effect on adsorption of LDLs \cite{Zhao2009, Dehghan2021b}, as well as other macromolecules \cite{Steinbach2013, Zhou2013, Wang2014, Kim2021, Aar2022, Badenhorst2024, Badenhorst2025, Mir2025}. We introduce the adsorption effectiveness, $Q$, as the product of the adsorption capacity per unit area, $c=q/A$, and binding energy per single LDL, $E_\mathrm{bind}$. In this way, $Q$ combines both adsorption capacity and strength in a single characteristic. At a constant LDL concentration, just enough to cover all adsorbent area, $Q$ is monitored in a wide range of azobenzenes surface densities, $\rho_a$. For all three grafting patterns, $Q$ decreases with the increase of the polymer length. It is explained by the fact that the increase of the adsorbent mass does not lead to an increase of the number of bound LDLs, as only the top layer of a brush functions as an adsorbing surface. Therefore, our further analysis is focused on the case of the shortest polymers with the backbone length of $l_\mathrm{bb}=6$ monomers. In this case, all three patterns display the sharp peak at $\rho_a\approx 0.1$, but its height for the grid-like and bunch-like patterns is larger by one-third and about twice, respectively, than the height for an uniform brush. Besides that, in the case of a bunch-like pattern, another broad and relatively broad peak for $Q$ is observed at about $\rho_a\approx 0.3$, whereas for two other patterns the values of $Q$ in this density interval are close to zero. This factor extends the usability of the bunch-like pattern to the region with higher azobenzene surface density.

To clarify the difference between the adsorption effectiveness, $Q$, at $\rho_a\approx 0.1$ for the bunch-like and uniform patterns, we performed the analysis of the partial contributions to $c$ and $E_\mathrm{bind}$, originated from different arrangement types of LDLs, $1-4$, with respect to the grid of grafting points. The cause of the sharp peak for $Q$ at $\rho_a\approx 0.12$ for a bunch-like pattern is the prevalence in this case of the arrangement of type $4$, when each LDL is located at the middle of a square formed by four grafting points. The peak at $\rho_a\approx 0.3$ is characterized by a more even presence of all arrangement types, $1-4$. Their partial contributions are summed up to form a second peak for $Q$. For the case of a uniform pattern, all partial contributions are approximately equal, as expected.

The adsorption isotherm is constructed at specific conditions, $l_\mathrm{bb}=6$, $\rho_a=0.12$, in which case the adsorption effectiveness of a bunch-like arrangement is the highest. The volume fraction of LDLs, $f_\mathrm{LDL}$, varies from $0$ to $0.18$. Adsorption isotherms exhibit the lower density ascending regime, and higher density descending regime. The former is fitted well by either the Langmuir or logistic growth forms, providing the adsorption constant and the growth rate, respectively. Both fits indicate higher values for these characteristics for the bunch-like pattern compared with the uniform pattern, the increase is from $33\%$ to $100\%$. On the contrary, the shape of $Q$, as the function of $f_\mathrm{LDL}$, has a wide plateau region at middle values of $0.05<f_\mathrm{LDL}<0.14$, with the values for $Q$ for the bunch-like pattern about twice larger than the ones for the uniform pattern. One may deduce that the adsorber is effective in a rather wide interval of LDL concentrations. The descending regime is explained by ordering of bulk LDL into a cubic phase, thus, hampering the adsorption.

The simulation setup used in this work could be improved in a number of ways in the future studies. In particular, the model for high density lipoproteins can be developed, and, consequently, the adsorption selectivity of an adsorbent towards LDL can be modelled in an explicit way. Besides that, here we focused on the trends in the adsorbent effectiveness upon changing its parameters. However, the model could be brought closer to real systems by performing parametrisation that is based on atomistic simulation data. Another factor, that can be brought into attention, is fluctuation of the LDL shape upon the change of environment. We hope to address these issues in the following studies.

\section*{Acknowledgements}

Jaroslav Ilnytskyi thanks the NRFU grant no. 2023.05/0019 for financial support. All authors thank the Armed Forces of Ukraine for their protection during the conduct of this research.

\bibliographystyle{cmpj}
\bibliography{LDL_ads_azobrush_pattern_new}

\providecommand{\noopsort}[1]{}
\begin{thebibliography}{10}
\providecommand{\url}[1]{\texttt{#1}}
\providecommand{\urlprefix}{URL }
\expandafter\ifx\csname urlstyle\endcsname\relax
  \providecommand{\doi}[1]{doi:\discretionary{}{}{}#1}\else
  \providecommand{\doi}{doi:\discretionary{}{}{}\begingroup
  \urlstyle{rm}\Url}\fi
\providecommand{\eprint}[2][]{\url{#2}}

\bibitem{Sacks2003}
Sacks~F.~M., Campos~H., J. Clin. Endocrinol. Metab., 2003, \textbf{88}, No.~10,
  4525--4532, \doi{10.1210/jc.2003-030636}.

\bibitem{Sikorski2007}
Sikorski~J., In: Comprehensive medicinal chemistry II, Elsevier, 2007,
  459--494, \doi{10.1016/b0-08-045044-x/00180-2}.

\bibitem{Mundi2017}
Mundi~S., Massaro~M., Scoditti~E., Carluccio~M.~A., {\noopsort{hinsbergh}}{van
  Hinsbergh}~V. W.~M., {Iruela-Arispe}~M.~L., Caterina~R.~D., Cardiovasc. Res.,
  2017, \textbf{114}, No.~1, 35--52, \doi{10.1093/cvr/cvx226}.

\bibitem{BasuRay2019}
BasuRay~S., Wang~Y., Smagris~E., Cohen~J.~C., Hobbs~H.~H., Proc. Natl. Acad.
  Sci. U.S.A., 2019, \textbf{116}, No.~19, 9521--9526,
  \doi{10.1073/pnas.1901974116}.

\bibitem{Thierer2019}
Thierer~J.~H., Ekker~S.~C., Farber~S.~A., Nat. Commun., 2019, \textbf{10},
  No.~1, 3426, \doi{10.1038/s41467-019-11259-w}.

\bibitem{Ronco2022}
Ronco~C., Bellomo~R., Crit. Care, 2022, \textbf{26}, No.~1, 135,
  \doi{10.1186/s13054-022-04009-w}.

\bibitem{Zhao2009}
Zhao~L., Sun~D., Liu~M., Carbohydr. Polym., 2009, \textbf{78}, No.~4, 828--832,
  \doi{10.1016/j.carbpol.2009.07.008}.

\bibitem{Gunkel2013}
Gunkel~G., Huck~W. T.~S., J. Am. Chem. Soc., 2013, \textbf{135}, No.~18,
  7047--7052, \doi{10.1021/ja402126t}.

\bibitem{Yu2021}
Yu~Y., Dong~J., Ma~B., Jiang~X., Guo~C., Liu~Z., Chai~Y., Wang~L., Sun~L.,
  Ou~L., Li~W., J. Mater. Chem. B, 2021, \textbf{9}, No.~32, 6364--6376,
  \doi{10.1039/d1tb01220g}.

\bibitem{Dehghan2021a}
Dehghan~R., Barzin~J., J. Membr. Sci., 2021, \textbf{620}, 118878,
  \doi{10.1016/j.memsci.2020.118878}.

\bibitem{Fang2024}
Fang~F., Zhao~H., Wang~R., Chen~Q., Wang~Q., Zhang~Q., Materials, 2024,
  \textbf{17}, No.~5, 988, \doi{10.3390/ma17050988}.

\bibitem{Cheng2003}
Cheng~Y., Wang~S., Yu~Y., Yuan~Y., Biomaterials, 2003, \textbf{24}, No.~13,
  2189--2194, \doi{10.1016/s0142-9612(03)00047-4}.

\bibitem{Yu2022}
Yu~Y., Ma~B., Jiang~X., Guo~C., Liu~Z., Li~N., Chai~Y., Wang~L., Du~Y.,
  Wang~B., Li~W., Ou~L., J. Mater. Chem. B, 2022, \textbf{10}, No.~25,
  4856--4866, \doi{10.1039/d2tb00291d}.

\bibitem{Ergun2025}
Ergun~C., Eskizengin~H., J. Biomed. Mater. Res. A, 2025, \textbf{113}, No.~3,
  e37883, \doi{10.1002/jbm.a.37883}.

\bibitem{Lanzalaco2017}
Lanzalaco~S., Armelin~E., Gels, 2017, \textbf{3}, No.~4, 36,
  \doi{10.3390/gels3040036}.

\bibitem{Sudre2020}
Sudre~G., Siband~E., Gallas~B., Cousin~F., Hourdet~D., Tran~Y., Polymers, 2020,
  \textbf{12}, No.~1, 153, \doi{10.3390/polym12010153}.

\bibitem{Yaremchuk2023}
Yaremchuk~D., Kalyuzhnyi~O., Ilnytskyi~J., Condens. Matter Phys., 2023,
  \textbf{26}, No.~3, 33302, \doi{10.5488/cmp.26.33302}.

\bibitem{Tian2023}
Tian~L., Dou~H., Shao~Y., Yi~Y., Fu~X., Zhao~J., Fan~Y., Ming~W., Ren~L., Chem.
  Eng. J., 2023, \textbf{456}, 141093, \doi{10.1016/j.cej.2022.141093}.

\bibitem{Malm2010}
Malm~J., Sahramo~E., Karppinen~M., Ras~R. H.~A., Chem. Mater., 2010,
  \textbf{22}, No.~11, 3349--3352, \doi{10.1021/cm903831c}.

\bibitem{Bahl2020}
Bahl~S., Nagar~H., Singh~I., Sehgal~S., Mater. Today Proc., 2020, \textbf{28},
  1302--1306, \doi{10.1016/j.matpr.2020.04.505}.

\bibitem{Guo2022}
Guo~C., Yu~Y., Jiang~X., Ma~B., Liu~Z., Chai~Y., Wang~L., Wang~B., Du~Y.,
  Li~N., Fan~H., Ou~L., ACS Appl. Mater. Interfaces, 2022, \textbf{14}, No.~30,
  34388--34399, \doi{10.1021/acsami.2c07193}.

\bibitem{Guo2024}
Guo~C., Jiang~X., Guo~X., Liu~Z., Wang~B., Du~Y., Tian~Z., Wang~Z., Ou~L.,
  Regen. Biomater., 2024, \textbf{860}, r253--262, \doi{10.1093/rb/rbae045}.

\bibitem{Wei2015}
Wei~Y.-b., Tang~Q., Gong~C.-b., Lam~M. H.-W., Anal. Chim. Acta, 2015,
  \textbf{900}, 10--20, \doi{10.1016/j.aca.2015.10.022}.

\bibitem{DeMartino2020}
De~Martino~S., Mauro~F., Netti~P.~A., Riv. Nuovo Cimento, 2020, \textbf{43},
  No.~12, 599--629, \doi{10.1007/s40766-021-00014-x}.

\bibitem{Merritt2021}
Merritt~I. C.~D., Jacquemin~D., Vacher~M., Phys. Chem. Chem. Phys., 2021,
  \textbf{23}, No.~35, 19155--19165, \doi{10.1039/d1cp01873f}.

\bibitem{Sandhu1986}
Sandhu~S., Yianni~Y., Morgan~C., Taylor~D., Zaba~B., Biochim. Biophys. Acta,
  Biomembr., 1986, \textbf{860}, No.~2, 253--262,
  \doi{10.1016/0005-2736(86)90521-3}.

\bibitem{Song1996}
Song~X., Geiger~C., Vaday~S., Perlstein~J., Whitten~D.~G., J. Photochem.
  Photobiol. Chem., 1996, \textbf{102}, No.~1, 39--45,
  \doi{10.1016/s1010-6030(96)04362-6}.

\bibitem{Delova2025}
Delova~A., Makhloutah~A., Bernhard~Y., Pasc~A., Monari~A., Chem. Eur. J., 2025,
  \textbf{31}, No.~69, e01824, \doi{10.1002/chem.202501824}.

\bibitem{Pritzl2025}
Pritzl~S.~D., Morstein~J., Pritzl~N.~A., Lipfert~J., Lohm{\"u}ller~T.,
  Trauner~D.~H., Commun. Mater., 2025, \textbf{6}, No.~1, 59,
  \doi{10.1038/s43246-025-00773-8}.

\bibitem{Tomoshige2025}
Tomoshige~S., Kawasaki~Y., Morimoto~J., Sato~N., Hashimoto~Y., Ishikawa~M.,
  Chem. Pharm. Bull., 2025, \textbf{73}, No.~6, 568--573,
  \doi{10.1248/cpb.c25-00252}.

\bibitem{Guinart2025}
Guinart~A., Qutbuddin~Y., Schwille~P., Feringa~B.~L., Nat. Rev. Chem., 2025,
  \textbf{10}, No.~1, 12--30, \doi{10.1038/s41570-025-00783-7}.

\bibitem{Ilnytskyi2023}
Ilnytskyi~J., Yaremchuk~D., Komarytsia~O., Processes, 2023, \textbf{11},
  No.~10, 2913, \doi{10.3390/pr11102913}.

\bibitem{Ilnytskyi2006}
Ilnytskyi~J., Saphiannikova~M., Neher~D., Condens. Matter Phys., 2006,
  \textbf{9}, No.~1, 87, \doi{10.5488/cmp.9.1.87}.

\bibitem{Ilnytskyi2011}
Ilnytskyi~J.~M., Neher~D., Saphiannikova~M., J. Chem. Phys., 2011,
  \textbf{135}, No.~4, 044901, \doi{10.1063/1.3614499}.

\bibitem{Ilnytskyi2015}
Ilnytskyi~J.~M., Saphiannikova~M., ChemPhysChem, 2015, \textbf{16}, No.~15,
  3180--3189, \doi{10.1002/cphc.201500500}.

\bibitem{Ilnytskyi2016p}
Ilnytskyi~J., Slyusarchuk~A., Saphiannikova~M., Math. Model. Comput., 2016,
  \textbf{3}, No.~1, 33--42, \doi{10.23939/mmc2016.01.033}.

\bibitem{Ilnytskyi2016m}
Ilnytskyi~J.~M., Slyusarchuk~A., Saphiannikova~M., Macromolecules, 2016,
  \textbf{49}, No.~23, 9272--9282, \doi{10.1021/acs.macromol.6b01871}.

\bibitem{Ilnytskyi2018}
Ilnytskyi~J.~M., Slyusarchuk~A., Soko{\l}owski~S., Soft Matter, 2018,
  \textbf{14}, No.~19, 3799--3810, \doi{10.1039/c8sm00356d}.

\bibitem{Ilnytskyi2019}
Ilnytskyi~J.~M., Toshchevikov~V., Saphiannikova~M., Soft Matter, 2019,
  \textbf{15}, No.~48, 9894--9908, \doi{10.1039/c9sm01853k}.

\bibitem{Slyusarchuk2020}
Slyusarchuk~A.~Y., Yaremchuk~D.~L., Ilnytskyi~J.~M., Math. Model. Comput.,
  2020, \textbf{7}, No.~2, 207--218, \doi{10.23939/mmc2020.02.207}.

\bibitem{Yaremchuk2022}
Yaremchuk~D., Patsahan~T., Ilnytskyi~J., Condens. Matter Phys., 2022,
  \textbf{25}, No.~3, 33601, \doi{10.5488/cmp.25.33601}.

\bibitem{Xue2011}
Xue~C., {Yonet-Tanyeri}~N., Brouette~N., Sferrazza~M., Braun~P.~V.,
  Leckband~D.~E., Langmuir, 2011, \textbf{27}, No.~14, 8810--8818,
  \doi{10.1021/la2001909}.

\bibitem{Dehghan2021b}
Dehghan~R., Barzin~J., Sep. Purif. Technol., 2021, \textbf{278}, 119512,
  \doi{10.1016/j.seppur.2021.119512}.

\bibitem{Steinbach2013}
Steinbach~A., Paust~T., Pluntke~M., Marti~O., Volkmer~D., ChemPhysChem, 2013,
  \textbf{14}, No.~15, 3523--3531, \doi{10.1002/cphc.201300516}.

\bibitem{Zhou2013}
Zhou~Z., Yu~P., Geller~H.~M., Ober~C.~K., Biomacromolecules, 2013, \textbf{14},
  No.~2, 529--537, \doi{10.1021/bm301785b}.

\bibitem{Wang2014}
Wang~X., Berger~R., Ramos~J.~I., Wang~T., Koynov~K., Liu~G., Butt~H.-J., Wu~S.,
  RSC Adv., 2014, \textbf{4}, No.~85, 45059--45064, \doi{10.1039/c4ra07623k}.

\bibitem{Kim2021}
Kim~Y., Laradji~A.~M., Sharma~S., Zhang~W., Yadavalli~N.~S., Xie~J., Popik~V.,
  Minko~S., Angew. Chem., 2021, \textbf{134}, No.~7, e202110990,
  \doi{10.1002/ange.202110990}.

\bibitem{Aar2022}
A{\c c}ari~I.~K., Sel~E., {\"O}zcan~I., Ate{\c s}~B., K{\"o}ytepe~S.,
  Thakur~V.~K., Adv. Colloid Interface Sci., 2022, \textbf{305}, 102694,
  \doi{10.1016/j.cis.2022.102694}.

\bibitem{Badenhorst2024}
Badenhorst~R., Makaev~S., Yaremchuk~D., Sajjan~Y., Sulimov~A., Reukov~V.~V.,
  Lavrik~N.~V., Ilnytskyi~J., Minko~S., Langmuir, 2024, \textbf{40}, No.~13,
  7008--7020, \doi{10.1021/acs.langmuir.4c00062}.

\bibitem{Badenhorst2025}
Badenhorst~R., Makaev~S.~V., Parker~M., Marunych~R., Reukov~V.,
  Bdzi{\'n}ska~A., Korchynskyi~O., Kalyuzhnyi~O., Yaremchuk~D., Ilnytskyi~J.,
  Patsahan~T., Minko~S., ACS Appl. Mater. Interfaces, 2025, \textbf{17},
  No.~35, 49193--49209, \doi{10.1021/acsami.5c08747}.

\bibitem{Mir2025}
Mir~M., Wilson~L.~D., Surf. Interfaces, 2025, \textbf{73}, 107545,
  \doi{10.1016/j.surfin.2025.107545}.

\bibitem{AdroherBentez2023}
{Adroher-Ben{\'i}tez}~I., Morozova~T.~I., Catalini~G., Garc{\'i}a~N.~A.,
  Barrat~J.-L., Luengo~G.~S., L{\'e}onforte~F., Macromolecules, 2023,
  \textbf{56}, No.~24, 10285--10295, \doi{10.1021/acs.macromol.3c01503}.

\bibitem{Ilnytskyi2025}
Ilnytskyi~J., Yaremchuk~D., Minko~S., Langmuir, 2025, \textbf{41}, No.~20,
  12731--12744, \doi{10.1021/acs.langmuir.5c00929}.

\bibitem{Ma2019}
Ma~S., Zhang~X., Yu~B., Zhou~F., NPG Asia Mater., 2019, \textbf{11}, No.~1, 24,
  \doi{10.1038/s41427-019-0121-2}.

\bibitem{Pyun2003}
Pyun~J., Kowalewski~T., Matyjaszewski~K., Macromol. Rapid Commun., 2003,
  \textbf{24}, No.~18, 1043--1059, \doi{10.1002/marc.200300078}.

\bibitem{Edmondson2004}
Edmondson~S., Osborne~V.~L., Huck~W. T.~S., Chem. Soc. Rev., 2004, \textbf{33},
  No.~1, 14, \doi{10.1039/b210143m}.

\bibitem{Ouchi2009}
Ouchi~M., Terashima~T., Sawamoto~M., Chem. Rev., 2009, \textbf{109}, No.~11,
  4963--5050, \doi{10.1021/cr900234b}.

\bibitem{Ahmad2011}
Ahmad~S.~A., Leggett~G.~J., Hucknall~A., Chilkoti~A., Biointerphases, 2011,
  \textbf{6}, No.~1, 8--15, \doi{10.1116/1.3553579}.

\bibitem{Johnson2010}
Johnson~J.~A., Lu~Y.~Y., Burts~A.~O., Xia~Y., Durrell~A.~C., Tirrell~D.~A.,
  Grubbs~R.~H., Macromolecules, 2010, \textbf{43}, No.~24, 10326--10335,
  \doi{10.1021/ma1021506}.

\bibitem{Olivier2013}
Olivier~A., Meyer~F., Raquez~J.-M., Dubois~P., e-Polymers, 2013, \textbf{13},
  No.~1, 009, \doi{10.1515/epoly-2013-0109}.

\bibitem{Zhao2022}
Zhao~H., Gao~H., Chen~T., Xie~L., Ma~Y., Sha~J., Eur. Polym. J., 2022,
  \textbf{179}, 111469, \doi{10.1016/j.eurpolymj.2022.111469}.

\bibitem{Welch2012}
Welch~M.~E., Xu~Y., Chen~H., Smith~N., Tague~M.~E., Abruna~H.~D., Baird~B.,
  Ober~C.~K., J. Photopolym. Sci. Technol., 2012, \textbf{25}, No.~1, 53--56,
  \doi{10.2494/photopolymer.25.53}.

\bibitem{Yu2023}
Yu~B., Chang~B., Loo~W., Dhuey~S., O'Reilly~P., Ashby~P., Connolly~M.,
  Tikhomirov~G., Zuckermann~R., Ruiz~R., ACS Nano, 2024, \textbf{18},
  7411--423, \doi{10.1021/acsnano.3c10204}.

\bibitem{Jeon1999}
Jeon~N.~L., Choi~I.~S., Whitesides~G.~M., Kim~N.~Y., Laibinis~P.~E., Harada~Y.,
  Finnie~K.~R., Girolami~G.~S., Nuzzo~R.~G., Appl. Phys. Lett., 1999,
  \textbf{75}, No.~26, 4201--4203, \doi{10.1063/1.125582}.

\bibitem{Austin1986}
Austin~M., Krauss~R., Lancet, 1986, \textbf{328}, No. 8507, 592--595,
  \doi{10.1016/s0140-6736(86)92425-6}.

\bibitem{Campos1992}
Campos~H., Genest~J.~J., Blijlevens~E., McNamara~J.~R., Jenner~J.~L.,
  Ordovas~J.~M., Wilson~P.~W., Schaefer~E.~J., Arterioscler. Thromb. J. Vasc.
  Biol., 1992, \textbf{12}, No.~2, 187--195, \doi{10.1161/01.atv.12.2.187}.

\bibitem{Austin1988}
Austin~M.~A., JAMA J. Am. Med. Assoc., 1988, \textbf{260}, No.~13, 1917,
  \doi{10.1001/jama.1988.03410130125037}.

\bibitem{Stampfer1996}
Stampfer~M.~J., JAMA J. Am. Med. Assoc., 1996, \textbf{276}, No.~11, 882,
  \doi{10.1001/jama.1996.03540110036029}.

\bibitem{Lamarche1997}
Lamarche~B., Tchernof~A., Moorjani~S., Cantin~B., Dagenais~G.~R., Lupien~P.~J.,
  Despr\'es~J.-P., Circulation, 1997, \textbf{95}, No.~1, 69--75,
  \doi{10.1161/01.cir.95.1.69}.

\bibitem{Scheffer1997}
Scheffer~P.~G., Bakker~S. J.~L., Heine~R.~J., Teerlink~T., Clin. Chem., 1997,
  \textbf{43}, No.~10, 1904--1912, \doi{10.1093/clinchem/43.10.1904}.

\bibitem{Segrest2001}
Segrest~J.~P., Jones~M.~K., Loof~H.~D., Dashti~N., J. Lipid Res., 2001,
  \textbf{42}, No.~9, 1346--1367, \doi{10.1016/s0022-2275(20)30267-4}.

\bibitem{VanAntwerpen1994}
Antwerpen~R.~V., Gilkey~J.~C., J. Lipid Res., 1994, \textbf{35}, No.~12,
  2223--2231, \doi{10.1016/s0022-2275(20)39928-4}.

\bibitem{Teerlink2004}
Teerlink~T., Scheffer~P.~G., Bakker~S.~J., Heine~R.~J., J. Lipid Res., 2004,
  \textbf{45}, No.~5, 954--966, \doi{10.1194/jlr.m300521-jlr200}.

\bibitem{Ren2009}
Ren~G., Rudenko~G., Ludtke~S.~J., Deisenhofer~J., Chiu~W., Pownall~H.~J., Proc.
  Natl. Acad. Sci. U.S.A., 2009, \textbf{107}, No.~3, 1059--1064,
  \doi{10.1073/pnas.0908004107}.

\bibitem{Kumar2011}
Kumar~V., Butcher~S.~J., {\"O}{\"o}rni~K., Engelhardt~P., Heikkonen~J.,
  Kaski~K., {Ala-Korpela}~M., Kovanen~P.~T., PLoS ONE, 2011, \textbf{6}, No.~5,
  e18841, \doi{10.1371/journal.pone.0018841}.

\bibitem{Laguerre2007}
Laguerre~M., Lecomte~J., Villeneuve~P., Prog. Lipid Res., 2007, \textbf{46},
  No.~5, 244--282, \doi{10.1016/j.plipres.2007.05.002}.

\bibitem{Hevonoja2000}
Hevonoja~T., Pentik{\"a}inen~M.~O., Hyv{\"o}nen~M.~T., Kovanen~P.~T.,
  {Ala-Korpela}~M., Biochim. Biophys. Acta, Mol. Cell. Biol. Lipids, 2000, ,
  No.~3, 189--210, \doi{10.1016/s1388-1981(00)00123-2}.

\bibitem{Prassl2008}
Prassl~R., Laggner~P., Eur. Biophys. J., 2008, \textbf{38}, No.~2, 145--158,
  \doi{10.1007/s00249-008-0368-y}.

\bibitem{Dai2023}
Dai~L., Li~S., Hao~Q., Zhou~R., Zhou~H., Lei~W., Kang~H., Wu~H., Li~Y., Ma~X.,
  Nanoscale Adv., 2023, \textbf{5}, No.~4, 1011--1022,
  \doi{10.1039/d2na00883a}.

\bibitem{Murtola2011}
Murtola~T., Vuorela~T.~A., Hyv{\"o}nen~M.~T., Marrink~S.-J., Karttunen~M.,
  Vattulainen~I., Soft Matter, 2011, \textbf{7}, No.~18, 8135,
  \doi{10.1039/c1sm05367a}.

\bibitem{Pan2008}
Pan~Y., Tang~X., Eur. Polym. J., 2008, \textbf{44}, No.~2, 408--414,
  \doi{10.1016/j.eurpolymj.2007.11.004}.

\bibitem{Han2015}
Han~G., Zhang~H., Chen~J., Sun~Q., Zhang~Y., Zhang~H., New J. Chem., 2015,
  \textbf{39}, No.~2, 1410--1420, \doi{10.1039/c4nj01658k}.

\bibitem{Liu2024}
Liu~D., Zhao~J., Ma~Y., Zhao~X., Shi~S., Li~S., Song~Q., Cheng~X., Zhang~W.,
  Polym. Chem., 2024, \textbf{15}, No.~15, 1469--1474,
  \doi{10.1039/d4py00099d}.

\bibitem{Kong2025}
Kong~Q., Zhang~J., Zhang~K., Wang~S., He~M., Guo~Y., Gu~J., Angew. Chem. Int.
  Ed., 2025, \textbf{64}, No.~37, e202512721, \doi{10.1002/anie.202512721}.

\bibitem{Hughes2005}
Hughes~Z.~E., Wilson~M.~R., Stimson~L.~M., Soft Matter, 2005, \textbf{1},
  No.~6, 436, \doi{10.1039/b511082c}.

\bibitem{Herrero2025}
Gil~Herrero~C., Duve~T., Thallmair~S., J. Chem. Theory Comput., 2025,
  \textbf{22}, No.~1, 708--22, \doi{10.1021/acs.jctc.5c01654}.

\bibitem{Kihara1951}
Kihara~T., J. Phys. Soc. Jpn., 1951, \textbf{6}, No.~5, 289--296,
  \doi{10.1143/jpsj.6.289}.

\bibitem{Kihara1963}
Kihara~T., In: Advances in Chemical Physics, Vol.~5, Prigogine~I.  (Ed.),
  Wiley, 1963, 147--188, \doi{10.1002/9780470143513.ch3}.

\bibitem{Lintuvuori2008}
Lintuvuori~J.~S., Wilson~M.~R., J Chem Phys, 2008, \textbf{128}, No.~4, J.
  Chem. Phys., \doi{10.1063/1.2825292}.

\bibitem{Groot1997}
Groot~R.~D., Warren~P.~B., J. Chem. Phys., 1997, \textbf{107}, No.~11,
  4423--4435, \doi{10.1063/1.474784}.

\bibitem{Wilson1997}
Wilson~M.~R., J. Chem. Phys., 1997, \textbf{107}, No.~20, 8654--8663,
  \doi{10.1063/1.475017}.

\bibitem{Frauenthal1979}
Frauenthal~J.~C., Introduction to Population Modeling, Birkh\"auser, Boston,
  MA, 1979, \doi{10.1007/978-1-4684-7322-3}.

\bibitem{Price2023}
Price~T., In: Ecology of a changed world, chap.~2, Oxford University Press, New
  York, 2023, 9--20, \doi{10.1093/oso/9780197564172.003.0002}.

\end{thebibliography}

\ukrainianpart

\title{Адсорбція ліпопротеїнів низької щільності на мікроструктурованій полімерній щітці: комп'ютерне моделювання}
\author{Я. Ільницький\refaddr{label1,label2},
	Д. Яремчук\refaddr{label1},
	О.Комариця\refaddr{label3}}
\addresses{\addr{label1} Інститут фізики конденсованих систем імені І. Р. Юхновського НАН України, 79011  Львів, вул.~Свєнціцького, 1, Україна
	\addr{label2} Інститут прикладної математики та фундаментальних наук, Національний університет «Львівська політехніка», вул. Митрополита Андрея, 5, 79013 Львів, Україна
	\addr{label3} Кафедра внутрішньої медицини №2, Львівський медичний університет імені Данила Галицького, вул.~Пекарська, 69, 79010 Львів, Україна
}

\makeukrtitle

\begin{abstract}
	\tolerance=3000%
	Останніми роками опубліковано низку результатів експериментальних досліджень щодо розробки фотовідновлюваних адсорбентів на основі полімерних щіток, які спрямовані на використанні при гемоперфузійній терапії. Механізм останньої полягає в селективній адсорбції ліпопротеїнів низької щільності (ЛПНЩ) у крові, яка відбувається за межами організму. У нашому попередньому дослідженні [\,J.Ilnytskyi \textit{et al., Processes}, 2023, \textbf{11}, 2913] ми розробили мезоскопічні моделі як для ЛПНЩ, так і для фотовідновлюваного адсорбента з однорідно прищепленими ланцюжками, і дослідили специфіку адсорбції залежно від довжини ланцюжків та густини прищеплення. У поточному дослідженні цей аналіз поширено на випадки пучкоподібнї та сітчастої мікроструктури прищеплення. Ми встановили, що пучкоподібна мікроструктура є найефективнішою і демонструє піки своєї адсорбційної ефективності при двох характерних густинах прищеплення. Пік при низькій густині є вдвічі вищим, ніж для випадку однорідно прищеплених ланцюгів. Пік високої густини, який виникає лише для пучкоподібної мікроструктури, розширює робочий діапазон густини прищеплення. Пік низької густини характеризується співимірністю розмірів ЛПНЩ та характерним просторовим кроком мікроструктури. Ізотерми адсорбції, побудовані за цих умов, добре узгоджуються як з ленгмюровою, так і з логістичною формою росту. В обох випадках пучкоподібна мікроструктура характеризується значно вищою швидкістю росту, ніж для однорідно прищеплених ланцюжків. При найвищій концентрації ЛПНЩ адсорбція ускладнюється самозбіркою ЛПНЩ у просторово-впорядковану кубічну фазу.
	\keywords ліпопротеїни, адсорбція, азобензен, молекулярна динаміка
\end{abstract}

\end{document}